\documentclass[a4paper,11pt]{article}

\usepackage[margin=1in]{geometry}

\usepackage{amsmath}
\usepackage{amssymb}
\usepackage[hidelinks]{hyperref}
\usepackage[labelfont=bf]{caption}
\usepackage{subcaption}
\usepackage{cite}
\usepackage{tikz}
\usepackage{ifthen}
\usepackage{bm}
\usepackage{authblk}
\usepackage{varwidth}

\title{\bf Quantitative and Interpretable Order Parameters \\ for Phase Transitions from Persistent Homology
}

\author[a]{Alex Cole\thanks{\href{mailto:a.e.cole@uva.nl}{\texttt{a.e.cole@uva.nl}}}}
\author[b]{Gregory J. Loges\thanks{\href{mailto:gloges@wisc.edu}{\texttt{gloges@wisc.edu}}}}
\author[b]{Gary Shiu\thanks{\href{mailto:shiu@physics.wisc.edu}{\texttt{shiu@physics.wisc.edu}}}}

\affil[a]{\protect\begin{varwidth}[t]{\linewidth}\protect\centering GRAPPA and ITFA, Institute of Physics, University of Amsterdam, \par Science Park 904, 1098 XH Amsterdam, the Netherlands\protect\end{varwidth}}

\affil[b]{\protect\begin{varwidth}[t]{\linewidth}\protect\centering Department of Physics, University of Wisconsin-Madison, \par 1150 University Ave, Madison, WI 53706, USA\protect\end{varwidth}}

\date{}

\begin{document}

\maketitle
\vspace{2em}
\begin{abstract}
    \noindent
    We apply modern methods in computational topology to the task of discovering and characterizing phase transitions. As illustrations, we apply our method to four two-dimensional lattice spin models: the Ising, square ice, XY, and fully-frustrated XY models. In particular, we use persistent homology, which computes the births and deaths of individual topological features as a coarse-graining scale or sublevel threshold is increased, to summarize multiscale and high-point correlations in a spin configuration. We employ vector representations of this information called persistence images to formulate and perform the statistical task of distinguishing phases. For the models we consider, a simple logistic regression on these images is sufficient to identify the phase transition. Interpretable order parameters are then read from the weights of the regression. This method suffices to identify magnetization, frustration, and vortex-antivortex structure as relevant features for phase transitions in our models. We also define ``persistence'' critical exponents and study how they are related to those critical exponents usually considered.
\end{abstract}
\vfill

\newpage
\tableofcontents

\section{Introduction}

Given an unknown condensed matter system sitting
in front of you, the zeroth order question you may ask is: what is its phase structure? With sufficient technical ability, one may vary the various coupling constants, external temperature, etc, and measure its ensuing equilibrium configurations. One way to understand the phase structure is to carefully search through the entire parameter space, and deduce for which parameter regimes the system looks similar (i.e., the system remains in the same phase). In doing so, one may occassionally encounter boundaries of the parameter space where some symmetry is broken or some specific heat diverges, indicating a new phase.
Having identified these phases, a natural next question is how to distinguish them in practice, i.e.\ what order parameters describe the various phase transitions. 
These questions are naturally phrased in the language of machine learning (ML).  Namely, the question ``how many phases are there'' is an exercise in \emph{unsupervised learning}, while the question ``how are different phases distinguished'' is an exercise in \emph{supervised learning}. Note that this is an exericise in distinguishing statistical ensembles, and incurs some amount of uncertainty.

Recently, ML techniques have been applied to these very tasks. Unsupervised methods such as Principal Component Analysis, clustering algorithms ,and autoencoders have been used to identify phase transitions (see, e.g., \cite{Wang_2016,van_Nieuwenburg_2017,Wetzel_2017,Hu_2017,Woloshyn:2019oww}). Support vector machines have been shown to be a useful tool in quantifying characteristics of the Ising phase transition \cite{Giannetti_2019}. Supervised learning with neural networks has proven useful in this classification task (see, e.g., \cite{Carrasquilla_2017,Ch_ng_2017,Huembeli2017:1710.08382v2,Tanaka2016:1609.09087v2,Carrasquilla_2020}), but often lacks the desired level of interpretability.

In this paper we use \emph{persistent homology} \cite{edelsbrunner2000topological,zomorodian2005computing} (see \cite{zomorodian2005topology,edelsbrunner2010computational,murugan2019introduction} for reviews) as a tool for detecting and characterizing phase transitions. As illustrations, we apply our method to study two-dimensional lattice spin systems. Persistent homology is a technique from Topological Data Analysis (TDA) that identifies the births and deaths of topological features throughout a family of discrete complexes. This family often corresponds to the data set at various coarse-graining scales.
By now, persistent homology has been fruitfully applied in a wide variety of fields, including sensor networks \cite{desilva2007}, image processing \cite{carlsson2008local}, genomics \cite{chan2013topology}, protein structure \cite{Xia_2014,Gameiro2015}, neuroscience \cite{2018arXiv180605167S}, cosmology \cite{Cole:2017kve,Biagetti:2020skr} and string theory \cite{Cirafici_2016,Cole:2018emh}, to name only a few.
In the context of spin systems, persistent homology encodes multiscale and high-order correlations in a data set. The main takeaway from our work is that this representation of a spin system configuration is not only sufficient to distinguish phases in spin systems, but additionally provides interpretable order parameters for the phase transitions. For example, we find that persistent homology identifies such varied phenomena as magnetization, frustration, and (anti)vortices in spin systems. Additionally, as a multiscale technique, persistent homology can capture a system's approach towards scale-invariance, i.e.\ its critical behavior. We work with \emph{persistence images} \cite{adams2015persistence}, which are vectorized representations of persistent homology information. This framework allows us to define quantitative order parameters and quantify the uncertainty that a particular spin configuration belongs to a particular phase.

{Persistent topological methods have been applied to statistical mechanics in a few cases, but so far these applications have been largely qualitative in terms of statistics. Ref.~\cite{donato2016persistent} studied the relationship between phase transitions and topology changes in configuration space. More recently, \cite{2020arXiv200102616S} studied the relaxation dynamics of a two-dimensional Bose gas with persistent homology, and \cite{tran2020topological} performed unsupervised learning on persistence diagrams to visualize their phase structure. However, the use of persistent topological methods in obtaining {\it quantitative} information about statistical mechanics systems is in our view an underdeveloped subject. The purpose of this paper is to provide an initial foray in this important direction.}
 
In this manuscript we use persistent homology to quantitatively characterize phase transitions in four different lattice spin systems. We consider discrete and continuous spin models with and without frustration in the ground state: see Figure~\ref{fig:model_overview}. Each example contains a distinct lesson. We begin with an obligatory analysis of the two-dimensional Ising model (Is). We are able to easily identify the model's phase transition relying only on training data far from the critical temperature. The magnetization as order parameter is immediately extracted from the weights of the corresponding logistic regression. Additionally, we examine the multiscale nature of the information probed by persistence images. In particular, we define persistence critical exponents that capture the model's approach towards criticality, finding interesting connections to the critical exponents usually considered. We then turn to the square-ice model (SI), for which there is no local order parameter due to frustrated low-energy dynamics. We again find a successful classification and are able to identify an order parameter associated with the low-temperature phase's ``scale of frustration.''  Our technique quickly picks up on the scale of this feature. We subsequently turn to continuous spin models where a configuration consists of a map $f$ from the lattice $\Lambda$ to the unit circle $S^1$. For these models, rather than a coarse-graining scale we study the topologies of the sublevel sets $f^{-1}(-\pi,\nu]\subseteq \Lambda$. We study the XY model, where a simple logistic regression on the persistence images discovers the Kosterlitz-Thouless (KT) phase transition and corresponding vortex-antivortex structure in the low-temperature phase. Vortex-antivortex pairs are shown to give a distinctive signature in the persistence images that the logistic regression discovers and decides to use on its own. Finally, we consider the fully-frustrated XY model (FFXY), where frustration prevents the formation of (anti)vortices in the low-temperature phase. In this case, our method identifies small scale correlations between next-to-nearest neighbors that reflect the system's attempt to satisfy competing constraints.

\begin{figure}[t]
    \centering
    \begin{tikzpicture}
        \newcommand{\x}{2.2}
        \newcommand{\y}{1.2}
        \draw (-2*\x,-2*\y) rectangle (2*\x,2*\y);
        \draw (0,-2*\y) -- (0,2*\y);
        \draw (-2*\x,0) -- (2*\x,0);
        \node[align=center] at (-\x,\y) {Ising\\ (Section~\ref{sec:ising})};
        \node[align=center] at (\x,\y) {Square-ice\\ (Section~\ref{sec:sqIce})};
        \node[align=center] at (-\x,-\y) {XY\\ (Section~\ref{sec:xy})};
        \node[align=center] at (\x,-\y) {Fully-frustrated XY\\ (Section~\ref{sec:ffxy})};
        \node[above] at (-\x,2*\y) {\bf Unfrustrated};
        \node[above] at (\x,2*\y) {\bf Frustrated};
        \node[above,rotate=90] at (-2*\x,\y) {\bf Discrete};
        \node[above,rotate=90] at (-2*\x,-\y) {\bf Continuous};
    \end{tikzpicture}
    \caption{Overview of models.}
    \label{fig:model_overview}
\end{figure}
 
An important feature of our analysis is the simplicity of our machine learning architecture. Once the relevant spin configurations are reduced to persistence images, the phase classification and extraction of order parameters can be achieved via a simple logistic regression. This reflects the fact that persistent homology condenses these data sets into their most relevant (and interpretable) features. 

The organization of this manuscript is as follows. In Section~\ref{sec:persistent_homology} we give a brief introduction to persistent homology, persistence images, and our computational choices for applying these techniques to spin models. In Section~\ref{sec:examples} we apply our methods to spin models of increasing complexity. We conclude in Section~\ref{sec:disc}.

\medskip
\paragraph{Note:}
While this manuscript was being prepared, \cite{olsthoorn2020finding} appeared, which considers the task of understanding the phase structure of a lattice model using persistent homology. Our work differs not only in the models considered (we study models with discrete as well as continuous spins) but also in the methods and goals.
\cite{olsthoorn2020finding} computes pairwise distances under a particular metric on the persistence diagrams of spin configurations and visualizes the phase diagram via a dimensional reduction. In this work, we consider the statistical task of distinguishing different phases from data (including quantitative uncertainties), which leads us to compute logistic regressions of persistence images. This approach enables us to 
obtain quantitative information of the statistical mechanical systems (such as the critical temperature and critical exponents), and provide physical interpretations of the order parameters.

\section{Persistent homology \& persistence images}
\label{sec:persistent_homology}
We are interested in developing general interpretable order parameters for phase transitions in spin systems. Some inspiration can be drawn from the hallowed two-dimensional (ferromagnetic) Ising model. In this case, spontaneous magnetization in the low-temperature phase leads to large, continuous domains where all spins are aligned. As the temperature is decreased towards $T=0$, these domains grow, so that at sufficiently low temperature, the entire system is aligned. On the other hand, in the high-temperature phase, spins receive enough thermal energy to be randomly oriented.
In the language of ML, the (non)existence and scale of magnetic domains manifests as a pattern in the hierarchical clustering of aligned spins. In other words, if we consider the set of aligned spins and perform successive coarse-graining transformations, we would be able to distinguish these two phases by the number of domains at different coarse-graining scales. Note that this is a \emph{multiscale} concept that probes high-order correlations functions.

In fact, clustering can be viewed as the most basic topological information about a data set, giving the total number of ``connected components.'' We may then consider the hierarchical (i.e.\ multiscale) topologies corresponding to higher-dimensional features as well: for example, loops. A unified description of topological features of all dimensions is given by algebraic topology, and the hierarchical or multiscale version of algebraic toplogy is persistent homology.

We now give a brief description of simplicial homology, referring the reader to  \cite{edelsbrunner2010computational,zomorodian2005topology} for details.
We begin by embedding our data in a discrete complex. 
We use both simplicial and cubical complexes in this work. In a simplicial complex, points (0-simplices) may be connected in pairs by edges (1-simplices), in triples by triangular faces (2-simplices), and so on. Simplicial complexes must be closed under taking faces: for example, if a 2-simplex is in the complex then so too must be its three edges and three vertices. A cubical complex is similar, but it consists of points (0-cubes), line segments (1-cubes), squares (2-cubes), and so on.
Topological aspects of the simplicial or cubical complex are then captured by its homology groups. These groups, denoted $H_p$, consist of equivalence classes of $p$-cycles, where two $p$-cycles are in the same equivalence class if they can be smoothly deformed into one another. $H_0$ consists of connected components, $H_1$ consists of noncontractible loops, and so on, with the Betti numbers $b_p$ giving the number of inequivalent, nontrivial $p$-cycles.

The core insight of \emph{persistent} homology is that such a procedure can be significantly enhanced in its stability and information context if instead of a single complex, a monotonically growing family, called a \emph{filtration}, is considered. Often the growing of the filtration corresponds to the increasing of a coarse-graining scale, so that multiscale information is captured. See Figure~\ref{fig:filtrationDemo}. As this coarse-graining scale increases, $p$-cycles are created (for example, loops form) and destroyed (for example, loops are ``filled in''). The mathematics of persistent homology allows us to track the births and deaths of \emph{individual} topological features. This information is usually summarized via a \emph{persistence diagram} (see Figure~\ref{fig:square_example}), which is a scatter plot of these births and deaths.

\begin{figure}[t]
    \centering
    \begin{tikzpicture}
        \newdimen\s
        \newdimen\dx
        \s = 0.8cm
        \dx = 4cm
        
        \draw[->, very thick, >=stealth] (2.75*\s,1.5*\s) -- (\dx-0.75\s,1.5*\s);
        \draw[->, very thick, >=stealth] (2.75*\s+\dx,1.5*\s) -- (2*\dx-0.75*\s,1.5*\s);
        \draw[->, very thick, >=stealth] (2.75*\s+2*\dx,1.5*\s) -- (3*\dx-0.75*\s,1.5*\s);
        
        \fill[red, opacity=0.3] (2*\dx,0) rectangle (2*\s+2*\dx,\s);
        \fill[red, opacity=0.3] (2*\dx,\s) -- (2*\dx,2*\s) -- (\s+2*\dx,\s);
        \fill[red, opacity=0.3] (2*\s+2*\dx,\s) -- (2*\s+2*\dx,2*\s) -- (\s+2*\dx,\s);
        \fill[red, opacity=0.3] (2*\dx,3*\s) -- (2*\dx,2*\s) -- (\s+2*\dx,3*\s);
        \fill[red, opacity=0.3] (2*\s+2*\dx,3*\s) -- (2*\s+2*\dx,2*\s) -- (\s+2*\dx,3*\s);
        \fill[black, opacity=0.3] (3*\dx,0) -- (3*\dx,2*\s) -- (\s+3*\dx,\s) -- (2*\s+3*\dx,2*\s) -- (2*\s+3*\dx,0) -- (3*\dx,0);
        \fill[black, opacity=0.3] (3*\dx,2*\s) -- (3*\dx,3*\s) -- (2*\s+3*\dx,3*\s) -- (2*\s+3*\dx,2*\s) -- (\s+3*\dx,3*\s) -- (3*\dx,2*\s);
        \fill[red, opacity=0.3] (3*\dx,2*\s) -- (\s+3*\dx,3*\s) -- (2*\s+3*\dx,2*\s) -- (\s+3*\dx,\s) -- (3*\dx,2*\s);
        \draw[red, very thick] (\dx,0) -- (2*\s+\dx,0) -- (2*\s+\dx,3*\s) -- (\dx,3*\s) -- (\dx,0);
        \draw[red, very thick] (\dx,\s) -- (2*\s+\dx,\s);
        \draw[red, very thick] (\s+\dx,0) -- (\s+\dx,\s);
        \draw[black, very thick] (2*\dx,0) -- (2*\dx,3*\s) -- (2*\s+2*\dx,3*\s) -- (2*\s+2*\dx,0) -- (2*\dx,0);
        \draw[red, very thick] (2*\dx,\s) -- (\s+2*\dx,0);
        \draw[red, very thick] (\s+2*\dx,\s) -- (2*\s+2*\dx,0);
        \draw[black, very thick] (2*\dx,\s) -- (2*\s+2*\dx,\s);
        \draw[red, very thick] (2*\dx,2*\s) -- (\s+2*\dx,3*\s) -- (2*\s+2*\dx,2*\s) -- (\s+2*\dx,\s) -- (2*\dx,2*\s);
        \draw[black, very thick] (\s+2*\dx,0) -- (\s+2*\dx,\s);
        \draw[black, very thick] (3*\dx,0) -- (3*\dx,3*\s) -- (2*\s+3*\dx,3*\s) -- (2*\s+3*\dx,0) -- (3*\dx,0);
        \draw[black, very thick] (3*\dx,\s) -- (\s+3*\dx,0) -- (\s+3*\dx,\s) -- (2*\s+3*\dx,0);
        \draw[black, very thick] (3*\dx,\s) -- (2*\s+3*\dx,\s);
        \draw[black, very thick] (3*\dx,2*\s) -- (\s+3*\dx,3*\s) -- (2*\s+3*\dx,2*\s) -- (\s+3*\dx,\s) -- (3*\dx,2*\s);
        \draw[red, very thick] (\s+3*\dx,\s) -- (\s+3*\dx,3*\s);
        \foreach \x in {0,...,2} {
        \foreach \y in {0,...,3} {
            \ifthenelse{\NOT \x = 1 \OR \NOT \y = 2}{
                \fill[red!70!black] (\s*\x, \s*\y) circle (0.07);
                \fill (\s*\x+\dx, \s*\y) circle (0.07);
                \fill (\s*\x+2*\dx, \s*\y) circle (0.07);
                \fill (\s*\x+3*\dx, \s*\y) circle (0.07);
            };
        }}
        
        \draw[thick] (\s,4.5*\s) circle (0.3*\s);
        \draw[thick] (\s,4.5*\s) -- (1.3*\s,4.5*\s);
        \draw[thick] (\s+\dx,4.5*\s) circle (0.5*\s);
        \draw[thick] (\s+\dx,4.5*\s) -- (1.5*\s+\dx,4.5*\s);
        \draw[thick] (\s+2*\dx,4.5*\s) circle (0.717*\s);
        \draw[thick] (\s+2*\dx,4.5*\s) -- (1.717*\s+2*\dx,4.5*\s);
        \draw[thick] (\s+3*\dx,4.5*\s) circle (\s);
        \draw[thick] (\s+3*\dx,4.5*\s) -- (2*\s+3*\dx,4.5*\s);
        \node at (1.5\s+3*\dx,4.7*\s) {\footnotesize$\alpha$};
    \end{tikzpicture}
    
    \vspace{15pt}
    \qquad\hrule\qquad
    \vspace{15pt}
    
    \begin{tikzpicture}
        \newcommand{\z}{0.3}
        \newcommand{\xshift}{3}
        
        \draw[dashed] (0.5*\xshift,-3.5*\z) -- (0.5*\xshift,3.5*\z);
        \draw[->, very thick, >=stealth] (4*\z+\xshift,0) -- (2*\xshift-4*\z,0);
        \draw[->, very thick, >=stealth] (4*\z+2*\xshift,0) -- (3*\xshift-4*\z,0);
        \draw[->, very thick, >=stealth] (4*\z+3*\xshift,0) -- (4*\xshift-4*\z,0);
        
        \draw[fill=black!10] (-3*\z,-4*\z) rectangle (-\z,-2*\z);
        \draw[fill=black!10] (-\z,-4*\z) rectangle (\z,-2*\z);
        \draw[fill=black!40] (\z,-4*\z) rectangle (3*\z,-2*\z);
        \draw[fill=black!40] (-3*\z,-2*\z) rectangle (-\z,0);
        \draw[fill=black!10] (-\z,-2*\z) rectangle (\z,0);
        \draw[fill=black!70] (\z,-2*\z) rectangle (3*\z,0);
        \draw[fill=black!70] (-3*\z,0) rectangle (-\z,2*\z);
        \draw[fill=black] (-\z,0) rectangle (\z,2*\z);
        \draw[fill=black!40] (\z,0) rectangle (3*\z,2*\z);
        \draw[fill=black!70] (-3*\z,2*\z) rectangle (-\z,4*\z);
        \draw[fill=black!40] (-\z,2*\z) rectangle (\z,4*\z);
        \draw[fill=black!10] (\z,2*\z) rectangle (3*\z,4*\z);
        
        \draw[fill=black!10] (-3*\z+\xshift,-4*\z) rectangle (-\z+\xshift,-2*\z);
        \draw[fill=black!10] (-\z+\xshift,-4*\z) rectangle (\z+\xshift,-2*\z);
        \draw[fill=black!10] (-\z+\xshift,-2*\z) rectangle (\z+\xshift,0);
        \draw[fill=black!10] (\z+\xshift,2*\z) rectangle (3*\z+\xshift,4*\z);
        \draw[red, very thick] (-2*\z+\xshift,-3*\z) -- (\xshift,-3*\z) -- (\xshift,-\z);
        \fill[red!70!black] (-2*\z+\xshift,-3*\z) circle (0.07);
        \fill[red!70!black] (\xshift,-3*\z) circle (0.07);
        \fill[red!70!black] (\xshift,-\z) circle (0.07);
        \fill[red!70!black] (2*\z+\xshift,3*\z) circle (0.07);
        
        \draw[fill=black!10] (-3*\z+2*\xshift,-4*\z) rectangle (-\z+2*\xshift,-2*\z);
        \draw[fill=black!10] (-\z+2*\xshift,-4*\z) rectangle (\z+2*\xshift,-2*\z);
        \draw[fill=black!40] (\z+2*\xshift,-4*\z) rectangle (3*\z+2*\xshift,-2*\z);
        \draw[fill=black!40] (-3*\z+2*\xshift,-2*\z) rectangle (-\z+2*\xshift,0);
        \draw[fill=black!10] (-\z+2*\xshift,-2*\z) rectangle (\z+2*\xshift,0);
        \draw[fill=black!40] (\z+2*\xshift,0) rectangle (3*\z+2*\xshift,2*\z);
        \draw[fill=black!40] (-\z+2*\xshift,2*\z) rectangle (\z+2*\xshift,4*\z);
        \draw[fill=black!10] (\z+2*\xshift,2*\z) rectangle (3*\z+2*\xshift,4*\z);
        \fill[red,opacity=0.3] (-2*\z+2*\xshift,-3*\z) rectangle (2*\xshift,-\z);
        \draw[red, very thick] (2*\z+2*\xshift,-3*\z) -- (-2*\z+2*\xshift,-3*\z) -- (-2*\z+2*\xshift,-\z) -- (2*\xshift,-\z) -- (2*\xshift,-3*\z);
        \draw[red, very thick] (2*\xshift,3*\z) -- (2*\z+2*\xshift,3*\z) -- (2*\z+2*\xshift,\z);
        \fill[red!70!black] (-2*\z+2*\xshift,-3*\z) circle (0.07);
        \fill[red!70!black] (2*\xshift,-3*\z) circle (0.07);
        \fill[red!70!black] (2*\xshift,-\z) circle (0.07);
        \fill[red!70!black] (2*\xshift,3*\z) circle (0.07);
        \fill[red!70!black] (-2*\z+2*\xshift,-\z) circle (0.07);
        \fill[red!70!black] (2*\z+2*\xshift,-3*\z) circle (0.07);
        \fill[red!70!black] (2*\z+2*\xshift,3*\z) circle (0.07);
        \fill[red!70!black] (2*\z+2*\xshift,\z) circle (0.07);
        
        \draw[fill=black!10] (-3*\z+3*\xshift,-4*\z) rectangle (-\z+3*\xshift,-2*\z);
        \draw[fill=black!10] (-\z+3*\xshift,-4*\z) rectangle (\z+3*\xshift,-2*\z);
        \draw[fill=black!40] (\z+3*\xshift,-4*\z) rectangle (3*\z+3*\xshift,-2*\z);
        \draw[fill=black!40] (-3*\z+3*\xshift,-2*\z) rectangle (-\z+3*\xshift,0);
        \draw[fill=black!10] (-\z+3*\xshift,-2*\z) rectangle (\z+3*\xshift,0);
        \draw[fill=black!70] (\z+3*\xshift,-2*\z) rectangle (3*\z+3*\xshift,0);
        \draw[fill=black!70] (-3*\z+3*\xshift,0) rectangle (-\z+3*\xshift,2*\z);
        \draw[fill=black!40] (\z+3*\xshift,0) rectangle (3*\z+3*\xshift,2*\z);
        \draw[fill=black!70] (-3*\z+3*\xshift,2*\z) rectangle (-\z+3*\xshift,4*\z);
        \draw[fill=black!40] (-\z+3*\xshift,2*\z) rectangle (\z+3*\xshift,4*\z);
        \draw[fill=black!10] (\z+3*\xshift,2*\z) rectangle (3*\z+3*\xshift,4*\z);
        \fill[red,opacity=0.3] (-2*\z+3*\xshift,-3*\z) rectangle (2*\z+3*\xshift,-\z);
        \draw[red, very thick] (-2*\z+3*\xshift,-\z) -- (-2*\z+3*\xshift,3*\z) -- (2*\z+3*\xshift,3*\z) -- (2*\z+3*\xshift,-3*\z) -- (-2*\z+3*\xshift,-3*\z) -- (-2*\z+3*\xshift,-\z) -- (2*\z+3*\xshift,-\z);
        \draw[red, very thick] (3*\xshift,-\z) -- (3*\xshift,-3*\z);
        \fill[red!70!black] (-2*\z+3*\xshift,-3*\z) circle (0.07);
        \fill[red!70!black] (3*\xshift,-3*\z) circle (0.07);
        \fill[red!70!black] (3*\xshift,-\z) circle (0.07);
        \fill[red!70!black] (3*\xshift,3*\z) circle (0.07);
        \fill[red!70!black] (-2*\z+3*\xshift,\z) circle (0.07);
        \fill[red!70!black] (-2*\z+3*\xshift,3*\z) circle (0.07);
        \fill[red!70!black] (2*\z+3*\xshift,-\z) circle (0.07);
        \fill[red!70!black] (-2*\z+3*\xshift,-\z) circle (0.07);
        \fill[red!70!black] (2*\z+3*\xshift,-3*\z) circle (0.07);
        \fill[red!70!black] (2*\z+3*\xshift,3*\z) circle (0.07);
        \fill[red!70!black] (2*\z+3*\xshift,\z) circle (0.07);
        
        \draw[fill=black!10] (-3*\z+4*\xshift,-4*\z) rectangle (-\z+4*\xshift,-2*\z);
        \draw[fill=black!10] (-\z+4*\xshift,-4*\z) rectangle (\z+4*\xshift,-2*\z);
        \draw[fill=black!40] (\z+4*\xshift,-4*\z) rectangle (3*\z+4*\xshift,-2*\z);
        \draw[fill=black!40] (-3*\z+4*\xshift,-2*\z) rectangle (-\z+4*\xshift,0);
        \draw[fill=black!10] (-\z+4*\xshift,-2*\z) rectangle (\z+4*\xshift,0);
        \draw[fill=black!70] (\z+4*\xshift,-2*\z) rectangle (3*\z+4*\xshift,0);
        \draw[fill=black!70] (-3*\z+4*\xshift,0) rectangle (-\z+4*\xshift,2*\z);
        \draw[fill=black] (-\z+4*\xshift,0) rectangle (\z+4*\xshift,2*\z);
        \draw[fill=black!40] (\z+4*\xshift,0) rectangle (3*\z+4*\xshift,2*\z);
        \draw[fill=black!70] (-3*\z+4*\xshift,2*\z) rectangle (-\z+4*\xshift,4*\z);
        \draw[fill=black!40] (-\z+4*\xshift,2*\z) rectangle (\z+4*\xshift,4*\z);
        \draw[fill=black!10] (\z+4*\xshift,2*\z) rectangle (3*\z+4*\xshift,4*\z);
        
        \fill[red, opacity=0.3] (-2*\z+4*\xshift,-3*\z) rectangle (2*\z+4*\xshift,3*\z);
        \draw[red, very thick] (-2*\z+4*\xshift,-3*\z) rectangle (2*\z+4*\xshift,3*\z);
        \draw[red, very thick] (-2*\z+4*\xshift,-\z) rectangle (2*\z+4*\xshift,\z);
        \draw[red, very thick] (4*\xshift,-3*\z) -- (4*\xshift,3*\z);
        \foreach\x in {-1,...,1} {
        \foreach\y in {0,...,3} {
            \fill[red!70!black] (2*\x*\z+4*\xshift,2*\y*\z-3*\z) circle (0.07);
        }}
    \end{tikzpicture}
    
    \vspace{10pt}
    \caption{(Top) Four steps in the $\alpha$-filtration for a grid of points, such as appears in our discrete-spin models. The filtration parameter when a $p$-simplex is included is $\alpha^2$, where $\alpha$ is the radius of the simplex's circumsphere. The $\alpha$-complexes are pictured in black/red, with the most recently added $p$-simplexes being shown in red. (Bottom) Four steps in the sublevel filtration for a function (represented by grayscale) defined on a $4\times3$ grid, such as appears in our continuous-spin models. The filtration parameter is the threshold, $\nu$, for the sublevel sets and the cubical complexes are pictured in red.}
    \label{fig:filtrationDemo}
\end{figure}

While persistence diagrams are often suitable for visualization, they are not very well suited for statistical analysis.
In the end, we are interested in the statistical task of quantifying the probability that a given spin configuration belongs to a particular phase of the system. Therefore, rather than scatter plots, we might prefer a summary statistic that lives in a vector space. These also aid us in quantitatively characterizing the change in the system's persistent homology as some parameter is varied.
We therefore use persistence images for our analysis, which are formed by appropriately smoothing the persistence diagram and binning so as to have a low(er)-dimensional representation of the persistence data. A weight factor which vanishes at zero persistence is used in order to highlight those more important features which have high persistence values. (See~\cite{adams2015persistence} for more details on the stability properties of persistence images.)

\subsection{Filtrations}
The data we consider come from square-lattice spin models, some with discrete, $\pm$, spins and others with continuous angles. We now describe the filtrations we use for these two cases, and give examples of how features of the spin configurations are captured by the persistence data.

\begin{figure}[t]
	\centering
	\begin{subfigure}{0.75\textwidth}
		\centering
		\includegraphics[width=\textwidth]{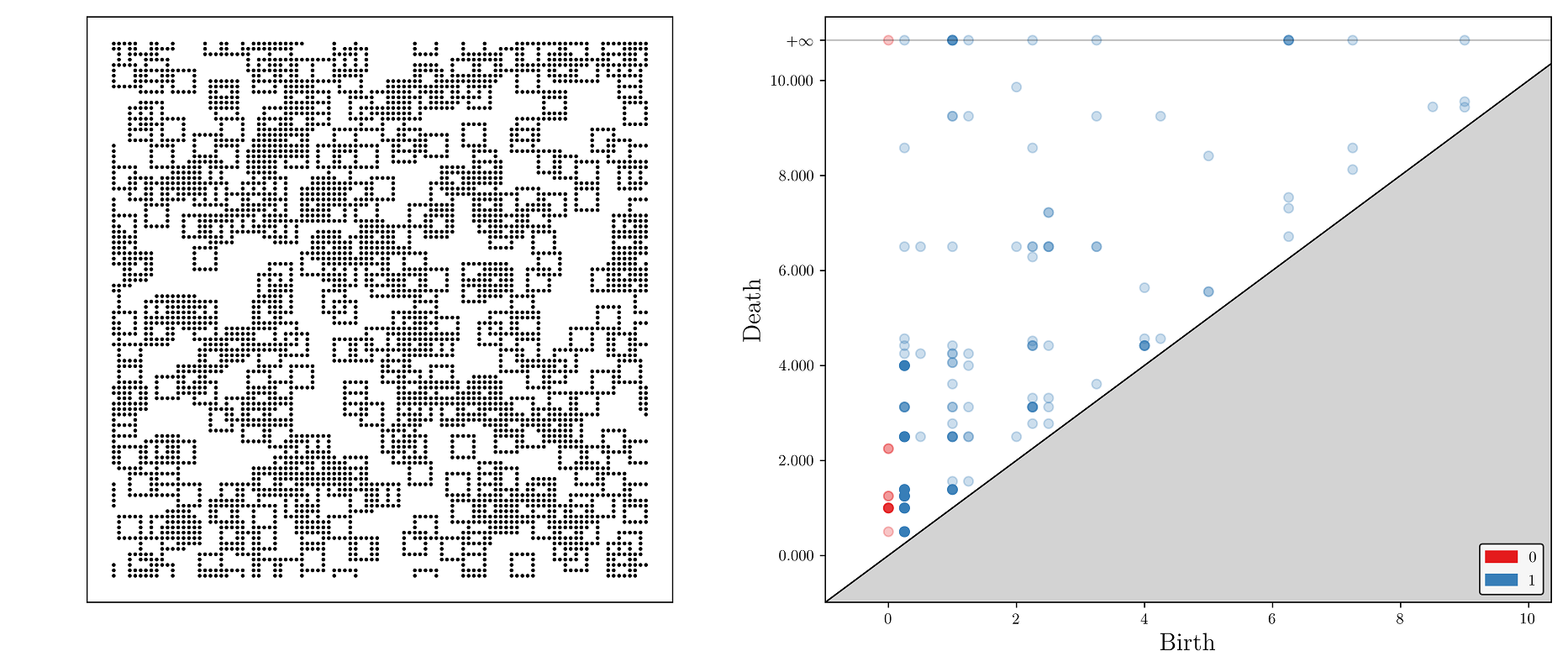}	
		\subcaption{Features of size $5\times5$.}
	\end{subfigure}\\
	\begin{subfigure}{0.75\textwidth}
		\centering
		\includegraphics[width=\textwidth]{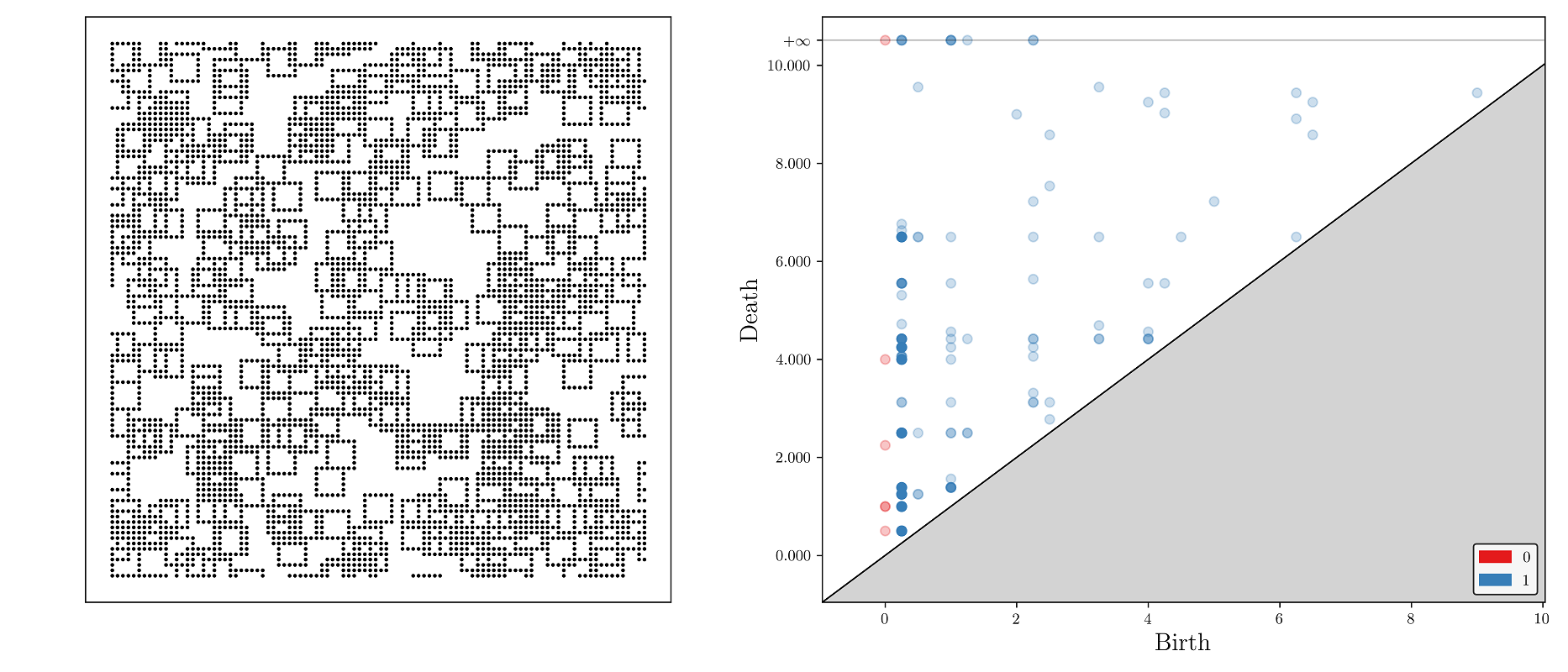}	
		\subcaption{Features of size $6\times6$.}
	\end{subfigure}
	\caption{Random configurations with features of a characteristic size and their corresponding persistence diagrams. Squares of spins are randomly flipped to be spin-up until 50\% of spins are aligned.}
	\label{fig:square_example}
\end{figure}

\paragraph{Filtrations for discrete spins}
With discrete spins on a square lattice we choose to represent our data via a point cloud, taking the locations of all spins aligned with a pre-determined direction as the data. We choose to take all spins which are aligned with the total magnetization (no matter how small). After creating the point cloud from a given spin configuration, we then use an $\alpha$-filtration to create the persistence diagram/image. The filtration corresponds to a coarse-graining of the point cloud, parameterized by the areas of balls enclosing the simplices: see~\ref{fig:filtrationDemo} for a small example.

As an example, Figure~\ref{fig:square_example} shows two (fake) spin configurations and their corresponding persistence diagrams. The configurations consist of (possibly overlapping) squares of aligned spins of a characteristic size. 1-cycles are born relatively early as adjacent, aligned spins in the lattice are connected by an edge. We see that there is a noticeable shift in the distribution of  persistences as the characteristic size of the squares changes. Note that this manifests in features of size smaller than the squares as well, due to their overlapping nature. Additionally, note that all 0-cycles are born at the beginning of the filtration. This is true in general for the $\alpha$-filtration. As such, when considering discrete spins we will generally use the 1-cycles for our statistical tests. To compute the persistence of $\alpha$-complexes we use the \texttt{GUDHI} class \texttt{AlphaComplex} \cite{gudhi:AlphaComplex}.

\begin{figure}[t]
	\centering
	\includegraphics[width=0.9\textwidth]{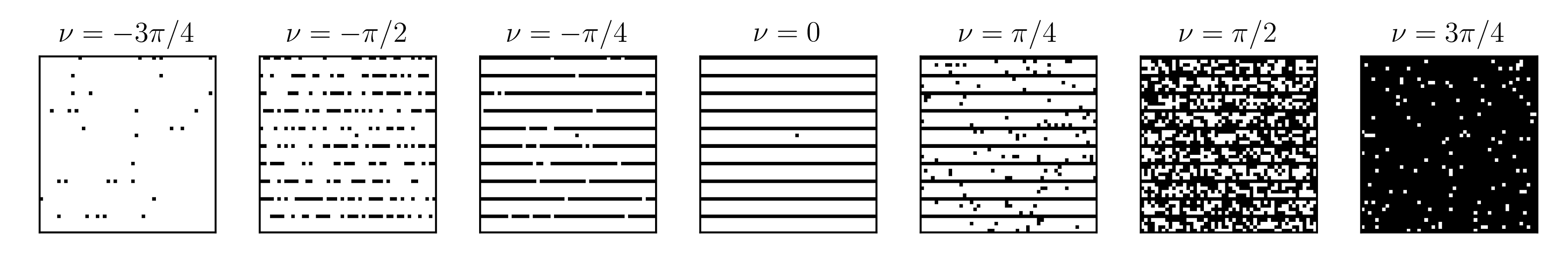}
	\includegraphics[width=0.9\textwidth]{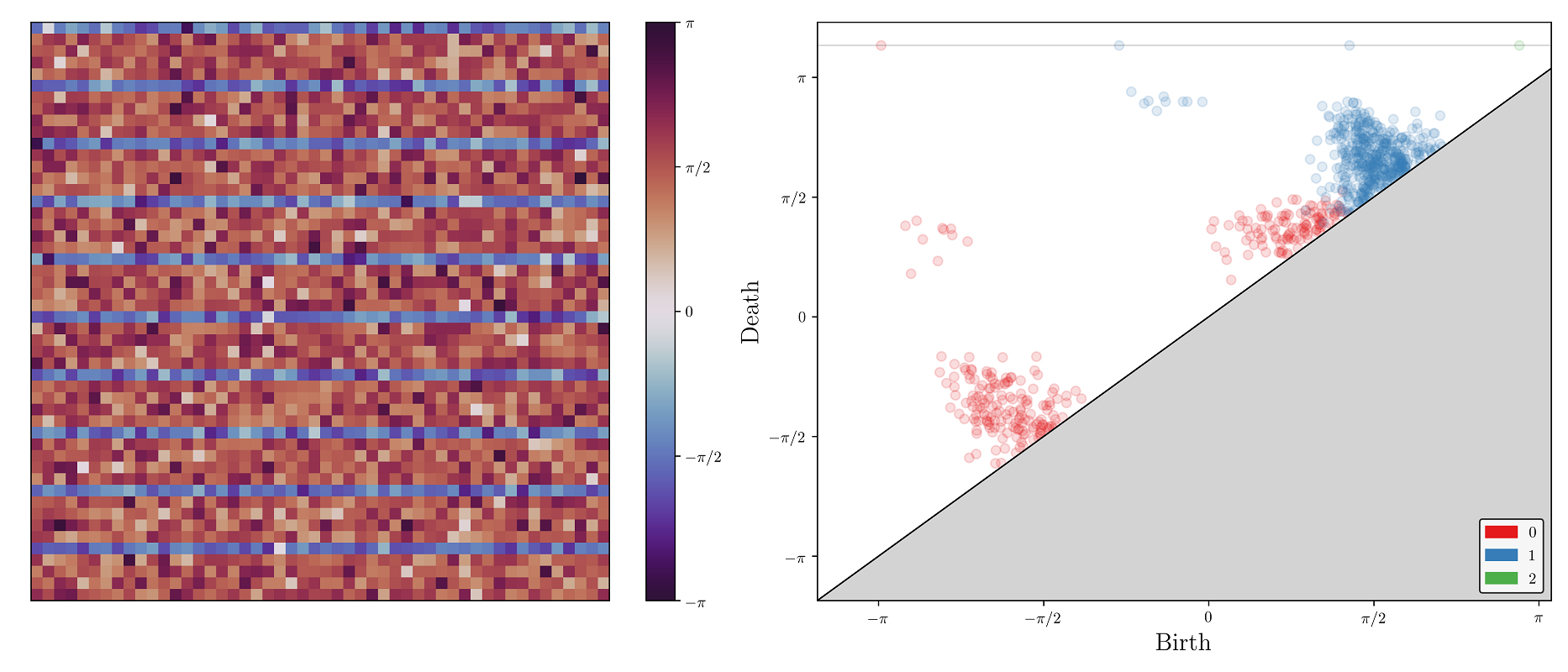}
	\caption{Random configuration with characteristic ``stripe'' pattern and its corresponding persistence diagram. Along the top we see steps in the sublevel filtration.}
	\label{fig:stripe_example}
\end{figure}

\paragraph{Filtrations for continuous spins}
We also consider models where the spins are continuous. In these cases, a spin configuration is a function $f:\Lambda\to S^1$ from the lattice, $\Lambda$, of $N$ spin sites to their angles. We consider models with global O(2) symmetry so that we may place the total magnetization (no matter how small) at angle $\theta=0$ and think of the function $f$ as mapping $\Lambda$ into $(-\pi,\pi]$. The sublevel sets with threshold $\nu\in S^1$, $f^{-1}(-\pi,\nu]$, then give a filtration of (periodic) cubical complexes. These sublevel sets experience topology change when the threshold $\nu$ passes a critical point of $f$, as is familiar from Morse theory \cite{milnor2016morse}. In this case 0-cycles have nontrivial births, corresponding to spin values where $f$ has a local minimum. As such we include both 0- and 1-cycles in the derived persistence images.

Figure~\ref{fig:stripe_example} shows an example continuous-spin configuration with a ``stripe'' pattern on top of Gaussian noise. There are a number of 0-cycles which are born early as the stripes appear in the filtration: the high-persistence 0-cycles are the 10 blue stripes which remain disconnected until joined by red spins. The high-persistence 1-cycles born near $\nu=0$ correspond to the nontrivial loops around the blue stripes (there are periodic boundary conditions). The low-persistence cycles are a result of the noise; for example for $\nu=\frac{3\pi}{4}$ there still remain a number of 1-cycles around those spins which just happen to have angles much larger than their neighbors. To compute the persistence of cubical complexes we use the \texttt{GUDHI} class \texttt{PeriodicCubicalComplex} \cite{gudhi:CubicalComplex}.

\begin{figure}[t]
	\centering
	\begin{subfigure}{0.22\textwidth}
		\centering
		\includegraphics[width=\textwidth]{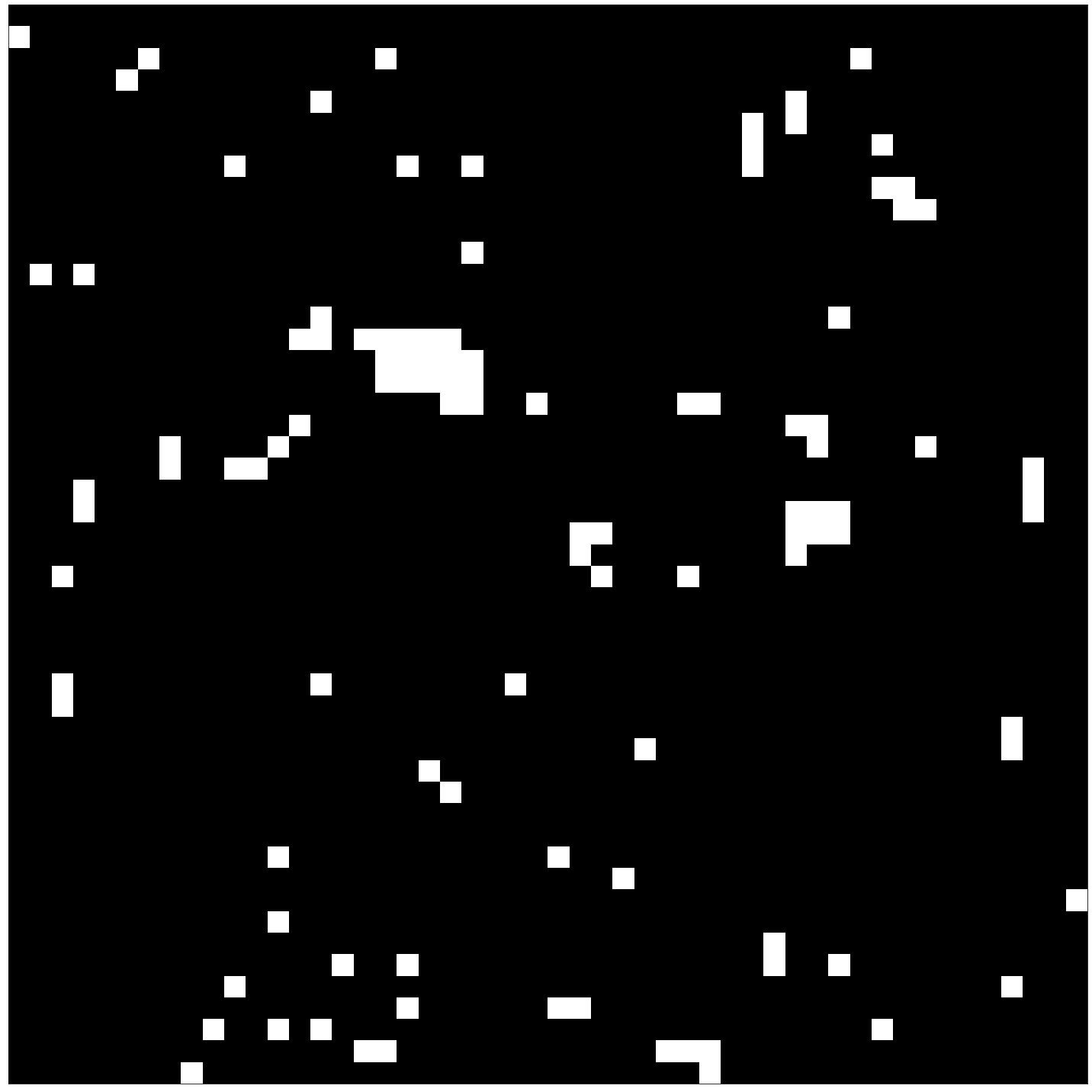}
		\subcaption{Is, $T=2.0$}
		\label{fig:Ising_low_temp}
	\end{subfigure}
	\begin{subfigure}{0.22\textwidth}
		\centering
		\includegraphics[width=\textwidth]{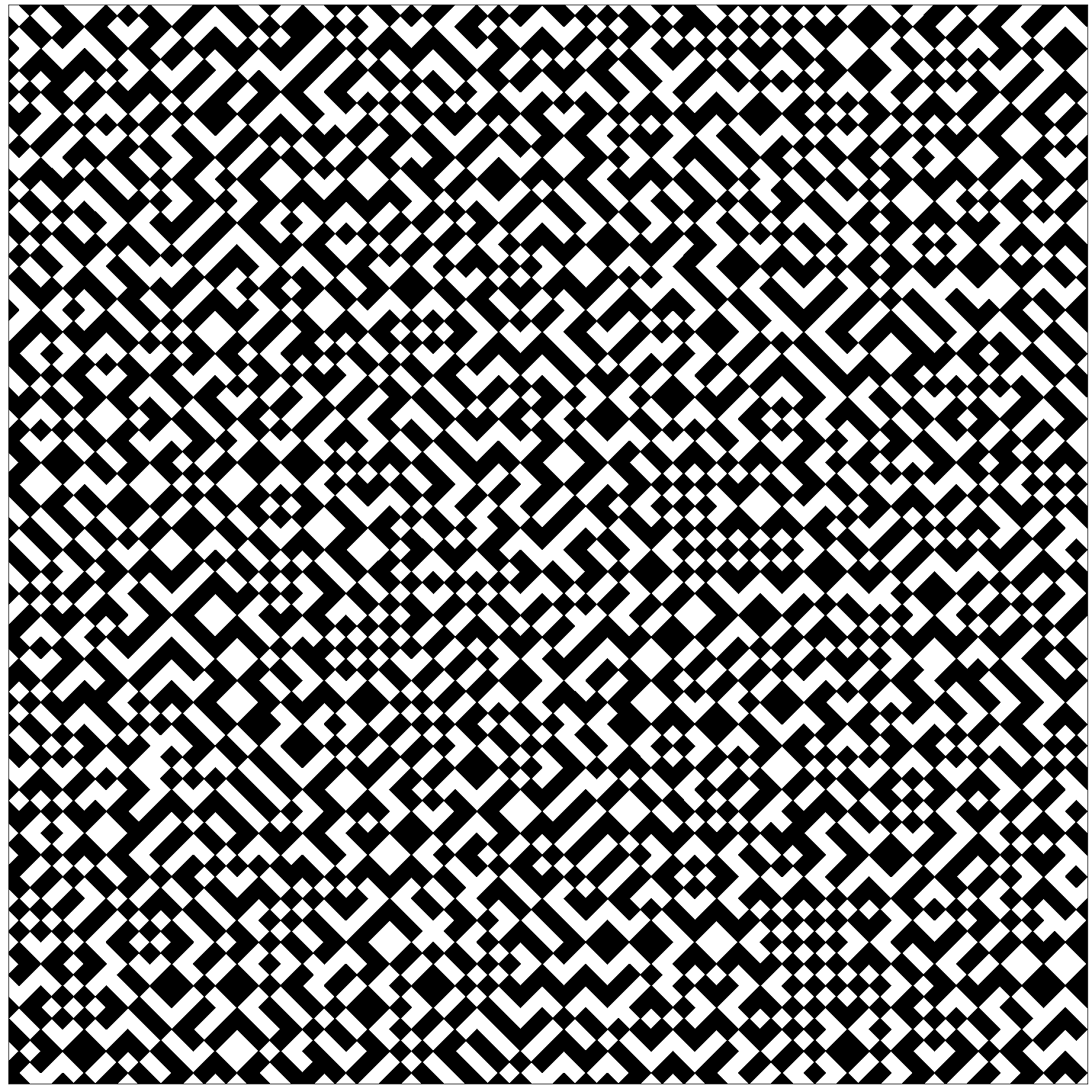}
		\subcaption{SI, $T=1.0$}
	\end{subfigure}
	\begin{subfigure}{0.22\textwidth}
		\centering
		\includegraphics[width=\textwidth]{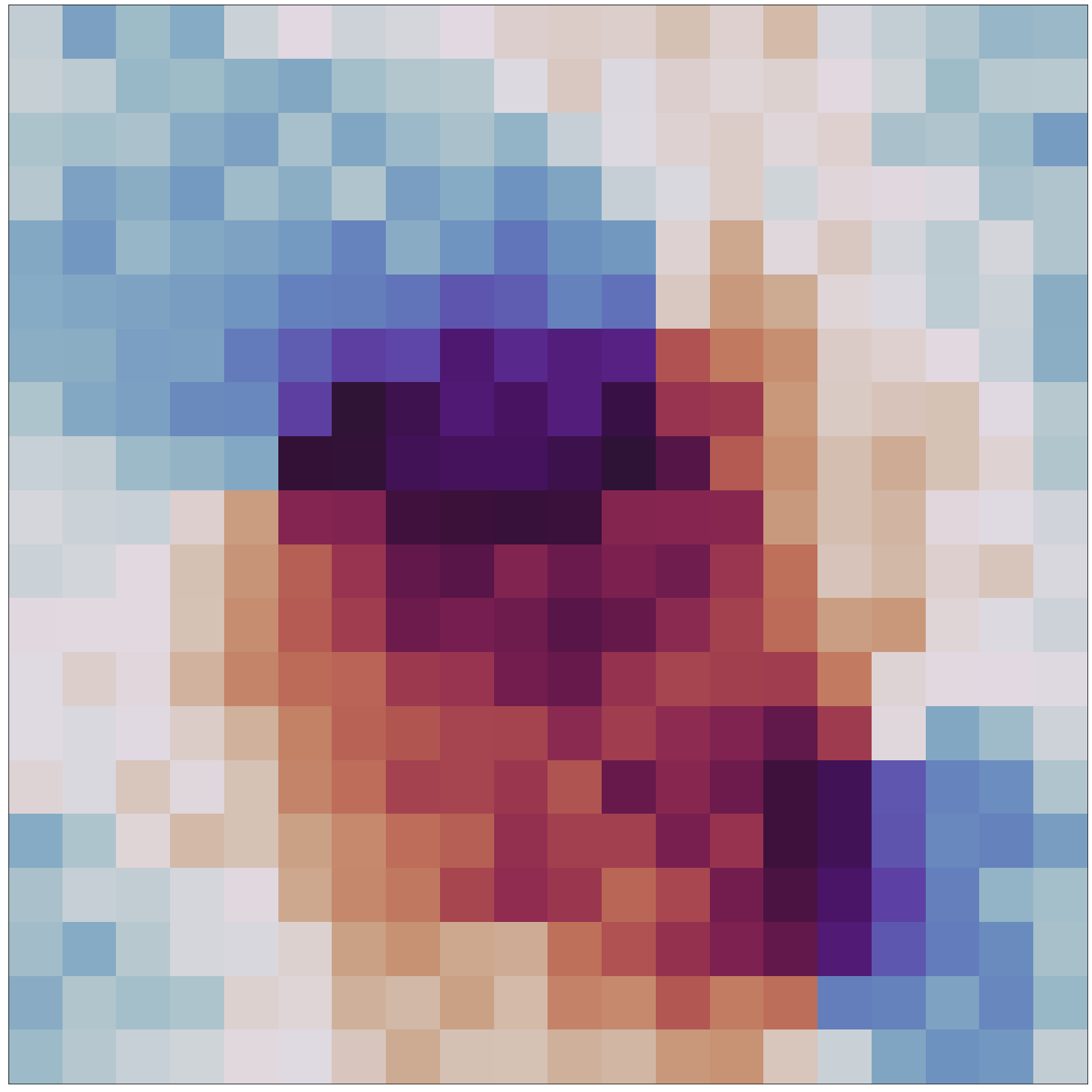}
		\subcaption{XY, $T=0.15$}
	\end{subfigure}
	\begin{subfigure}{0.22\textwidth}
		\centering
		\includegraphics[width=\textwidth]{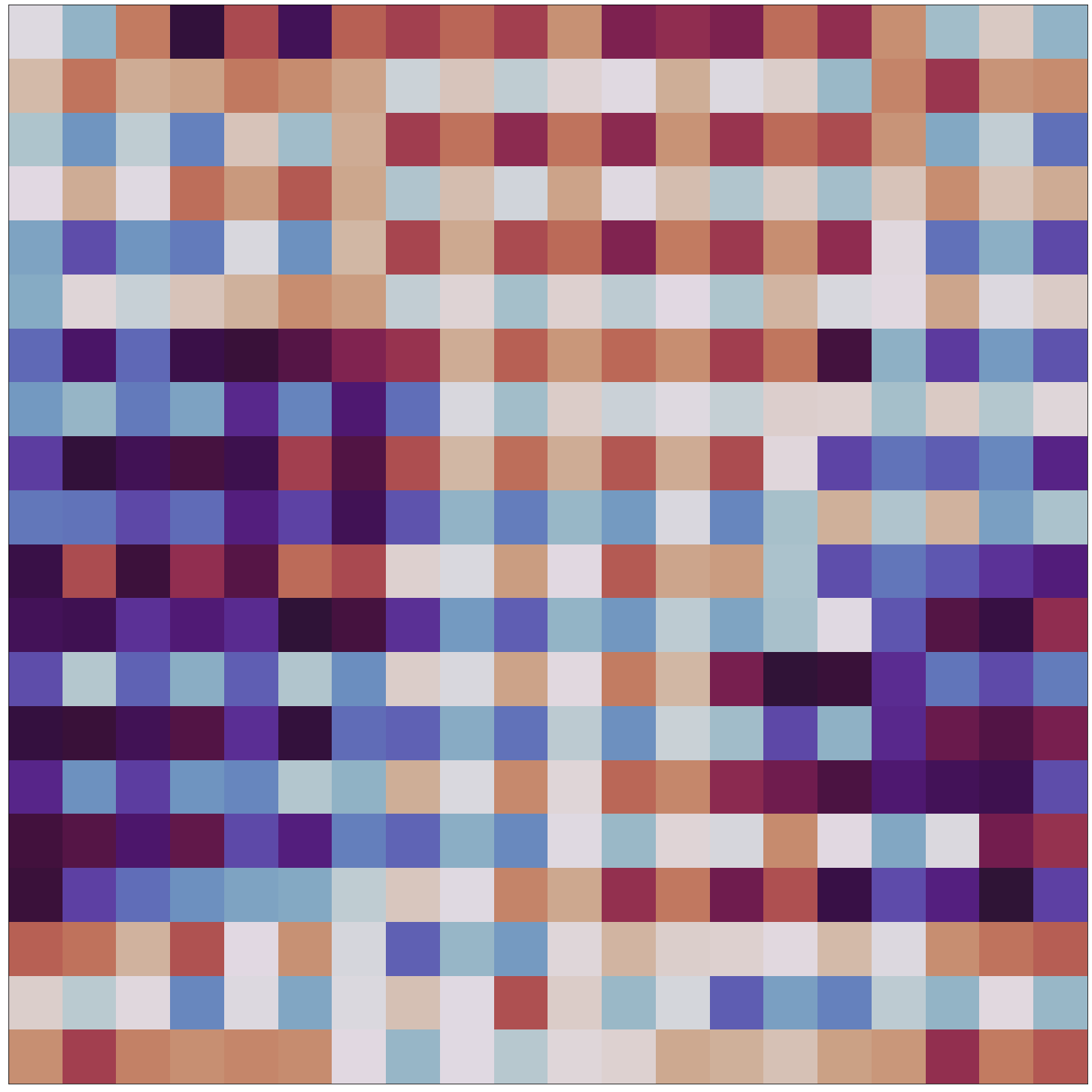}
		\subcaption{FFXY, $T=0.1$}
	\end{subfigure}\\[10pt]
	\begin{subfigure}{0.22\textwidth}
		\centering
		\includegraphics[width=\textwidth]{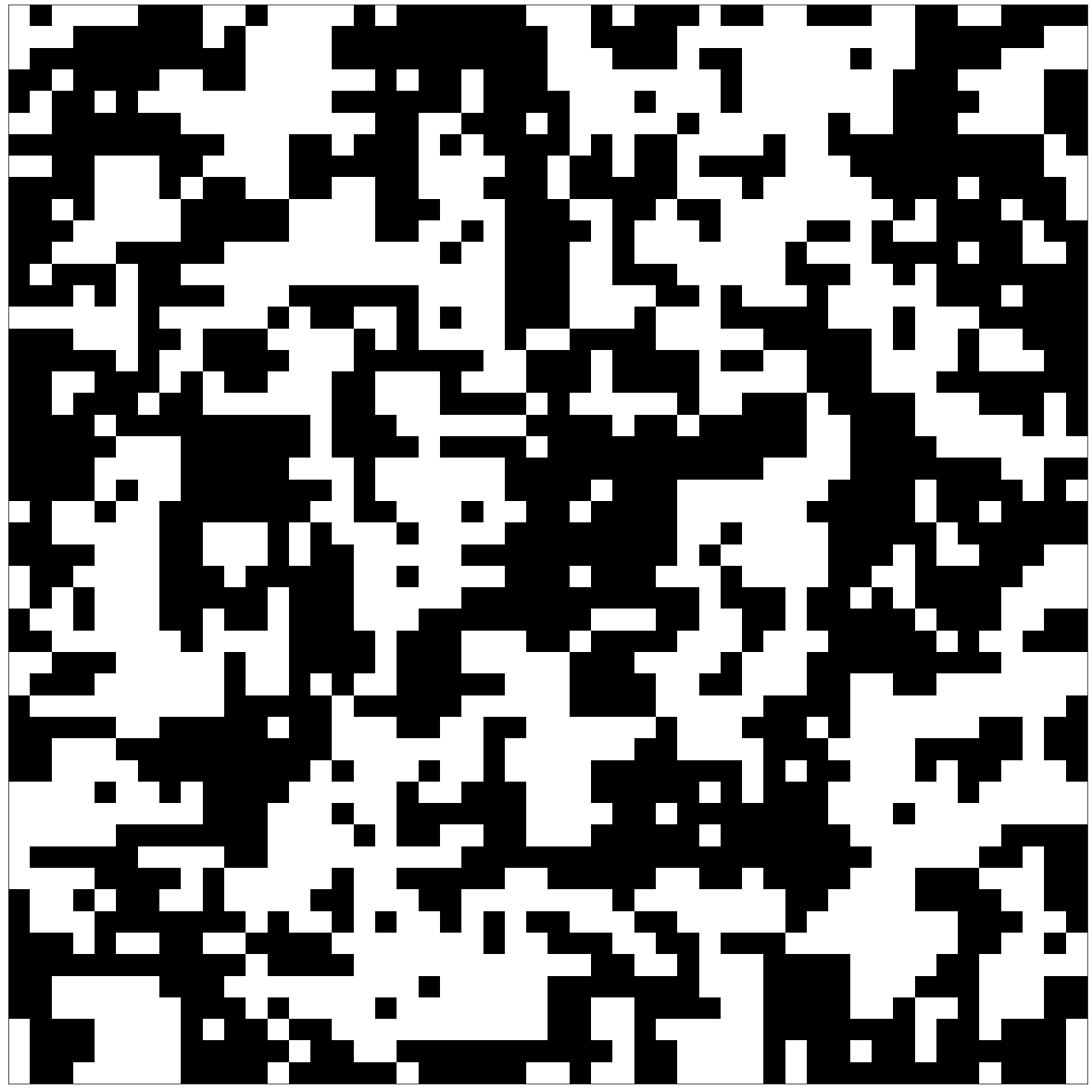}
		\subcaption{Is, $T=3.5$}
	\end{subfigure}
	\begin{subfigure}{0.22\textwidth}
		\centering
		\includegraphics[width=\textwidth]{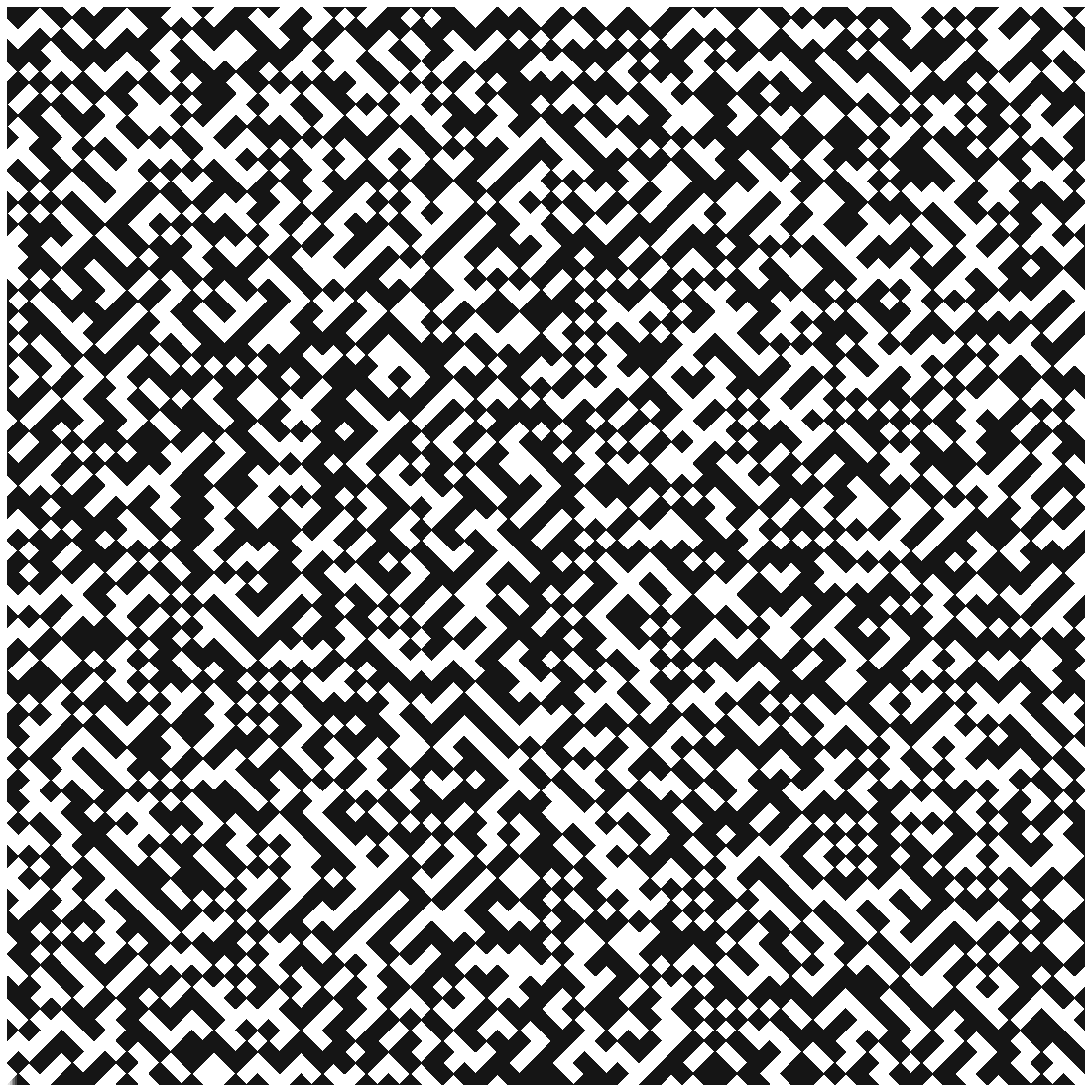}
		\subcaption{SI, $T=4.0$}
	\end{subfigure}
	\begin{subfigure}{0.22\textwidth}
		\centering
		\includegraphics[width=\textwidth]{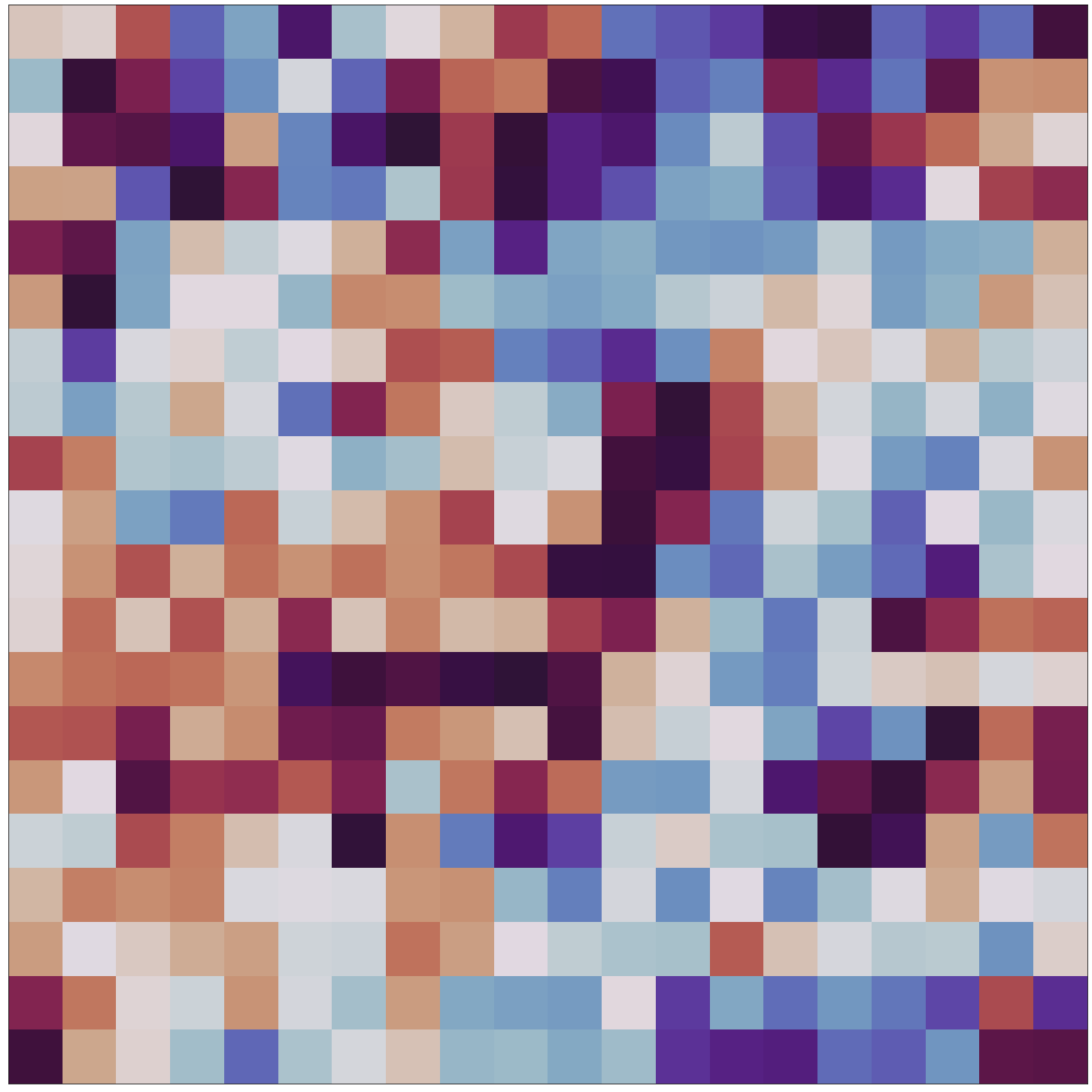}
		\subcaption{XY, $T=1.5$}
	\end{subfigure}
	\begin{subfigure}{0.22\textwidth}
		\centering
		\includegraphics[width=\textwidth]{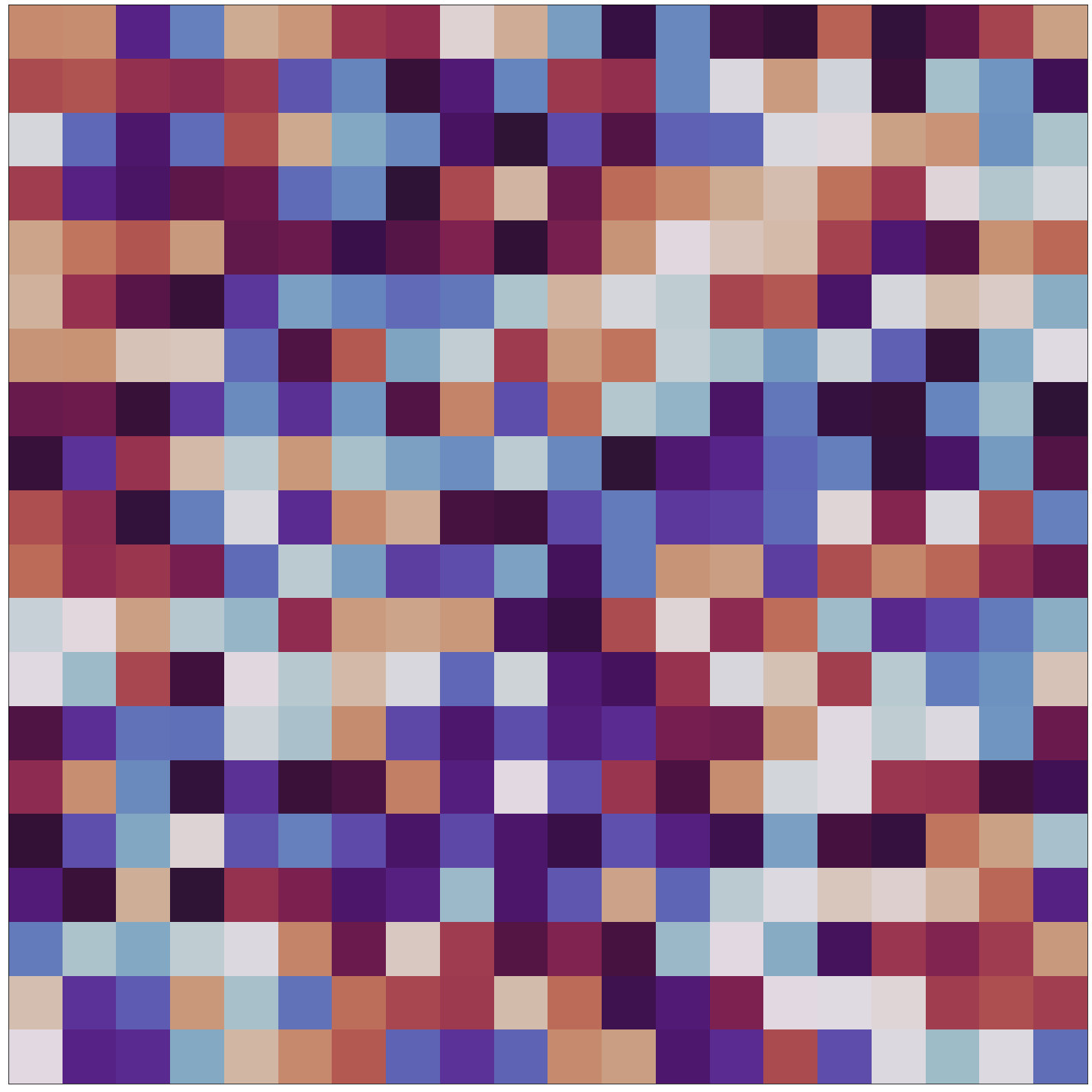}
		\subcaption{FFXY, $T=1.2$}
	\end{subfigure}
	\caption{Sample spin configurations in the low- and high-temperature phases for each model.}
	\label{fig:example_configs}
\end{figure}

\section{Phase classification \& critical phenomena}\label{sec:examples}

In this section we apply our methods to the task of phase classification in simple two-dimensional lattice spin models. We consider four such models: the Ising and square-ice models have discrete, $\pm$, spins and the XY and fully-frustrated XY models have continuously varying spins. Sample spin configurations for each model are generated at a number of temperatures using standard Monte-Carlo sampling techniques. Example spin configurations at low and high temperatures for each model are shown in Figure~\ref{fig:example_configs}.

For each model considered, classification into two phases is performed using only the persistence images. A subset of samples with extreme temperatures are used to train a logistic regression and then the accuracy of the regression is evaluated using the known temperatures of all samples. We normalize our persistence images using the $\ell_1$-norm, so they may be interpreted as probability densities for finding cycles with particular births/deaths for a given system. Unnormalized persistence images contain information about the total number of $p$-cycles and also lead to a successful classification.

\subsection{Logistic regression}
In the following sections we will be classifying spin configurations based on their persistence images. Since the persistence images are information-rich, we are able to use perhaps the simplest classification scheme, logistic regression, to great effect. Here we quickly recall the procedure of logistic regression. One benefit of logistic regression is that it is easy to tell what aspects of the data are used by the classification algorithm. We will use these to extract order parameters for the phase transitions under consideration.

Persistence images $\bm{x}\in\mathbb{R}^n$ ($n\sim\mathcal{O}(400)$ in our examples) are vectors of positive numbers representing the distribution of cycles in the birth-persistence plane. A logistic regression depends on parameters $\lambda_0$ and $\bm{\lambda}\in\mathbb{R}^n$ and the sigmoid function $\sigma:\mathbb{R}\to(0,1)$, given by
\begin{equation}
    \sigma(z) = \frac{1}{1+e^{-z}} \,.
\end{equation}
The sigmoid interpolates between $\sigma(-\infty)=0$ and $\sigma(\infty)=1$. A persistence image is declared to be in ``category~0'' if $\sigma(\lambda_0+\bm{\lambda}\cdot\bm{x})<\frac{1}{2}$ and in ``category~1'' if $\sigma(\lambda_0+\bm{\lambda}\cdot\bm{x})>\frac{1}{2}$. In our examples ``category~0'' will correspond to a low-temperature phase and ``category~1'' will correspond to a high-temperature phase. The parameters $\lambda_{i=0,\ldots,n}$ are learned by training on a subset of the data, $\bm{x}^{(k)}$, which are labeled into the two categories (i.e.~phases) with $y^{(k)}\in\{0,1\}$. Training amounts to maximize the log-likelihood, 
\begin{equation}
    \sum_k\left(y^{(k)}\log{\big[\sigma(\lambda_0+\bm{\lambda}\cdot\bm{x}^{(k)})\big]} + (1-y^{(k)})\log{\big[1-\sigma(\lambda_0+\bm{\lambda}\cdot\bm{x}^{(k)})\big]}\right) - C\sum_i\lambda_i^2 \,,
\end{equation}
where the constant $C$ controls the $\ell_2$-regularization used to prevent overfitting. By training on extreme temperatures, we incur some inaccuracy due to nonlinerarities as criticality is approached; these will not concern us too much, as we will find successful classification regardless.

Upon training, the regression can be applied to the rest of the persistence images to give an ``average classification'' at each temperature.  This can be interpreted as quantifying the regression's ``certainty'' that a particular temperature belongs to a particular phase. In addition, the learned coefficients $\lambda_i$ may be investigated to learn which bins (i.e.~regions) of the persistence images are most discerning when it comes to distinguishing the low- and high-temperature data. Bins where $\lambda_i\gg 0$ will identify features prevalent in the high-temperature phase, while $\lambda\ll 0$ will identify features that are prevalent in the low-temperature phase. These will constitute our order parameters.

\subsection{Ising model}\label{sec:ising}

\begin{figure}[!t]
	\centering
	\begin{subfigure}{0.9\textwidth}
		\centering
		\includegraphics[width=\textwidth]{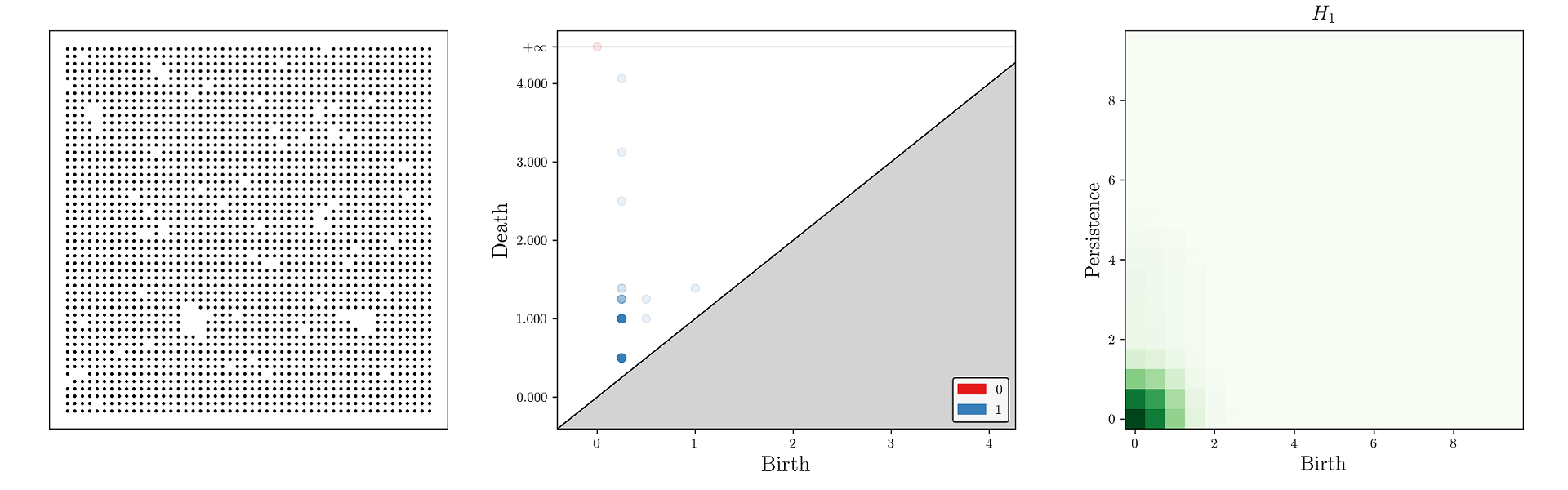}
		\includegraphics[width=\textwidth]{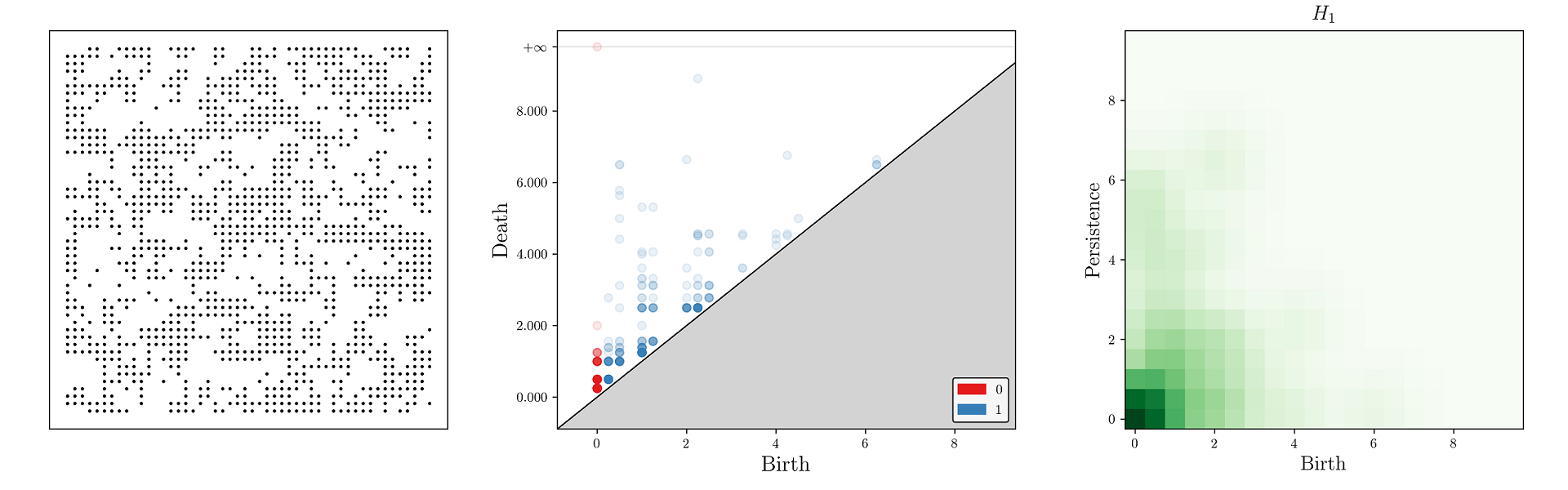}
		\subcaption{Sample spin configuration, persistence diagram and persistence image for $T=1.90$ (top) and $T=3.50$ (bottom).}
		\label{fig:Ising_PI}
	\end{subfigure}
	\begin{subfigure}{0.9\textwidth}
		\centering
		\includegraphics[width=\textwidth]{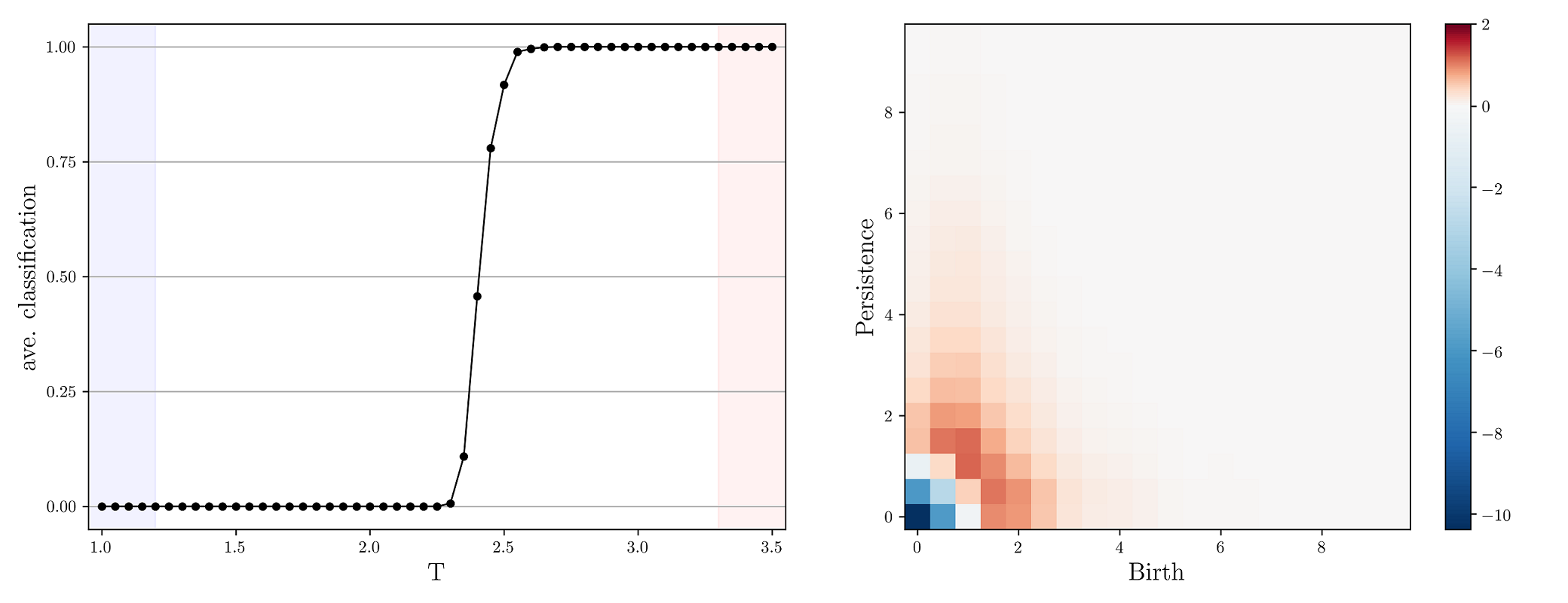}
		\subcaption{Average classification of testing data and learned logistic regression coefficients for the Ising model. The training data have temperatures in the highlighted regions. In the regression coefficients, blue regions are more populated in the low-temperature phase and red regions are more populated in the high-temperature phase.}
		\label{fig:Ising_log_reg}
	\end{subfigure}
	\caption{Ising model persistence data and phase classification.}
\end{figure}

The Ising model on a two-dimensional square lattice is very well understood, largely in part to Onsager's exact solution~\cite{PhysRev.65.117}. Spins $s_i\in\{{-1},1\}$ live at the vertices of the lattice with ferromagnetic interactions governed by the local Hamiltonian
\begin{equation}
	H_\text{Is} = -\sum_{\langle i,j\rangle}s_is_j \,,
\end{equation}
where the sum is over nearest-neighbor pairs. In the thermodynamic limit there is a second-order phase transition at $T_\text{Is} = \frac{2}{\log{(1+\sqrt{2})}}\approx 2.27$. At low temperatures there is spontaneous magnetization, while there is a disordered phase at high temperatures. While this model is well-understood, it provides a good first application of our method. We are able to easily extract the magnetization as order parameter from a simple logistic regression. We additionally study the relationship of new ``persistence'' critical exponents to those usually studied.

\subsubsection{Logistic regression and order parameter}
For temperatures $T\in\{1.00,1.05,\ldots,3.50\}$ we generate 1000 sample spin configurations for a $50\times50$ square lattice of 2500 spins. For each sample we construct the persistence image using a weight $\log{(1+p)}$, as in Figure~\ref{fig:Ising_PI}. Training of a logistic regression on the persistence images is conducted only on a subset of samples with extreme temperatures, well within the expected phases (see the left-hand side of Figure~\ref{fig:Ising_log_reg}). The classification extrapolates very well to the intermediate temperatures and gives an estimate of $T\approx 2.37$ for the critical temperature. The discrepancy from the known critical temperature may be attributed to finite-size effects.

The coefficients of the trained logistic regression (see the right-hand side of Figure~\ref{fig:Ising_log_reg}) show that the low-temperature configurations are identified by their having many small, short-lived cycles. These may be understood as arising both from $2\times 2$ blocks of aligned spins (which lead to very short-lived 1-cycles) as well as 1-cycles wrapping small groups of isolated spins which are flipped relative to the large domains of aligned spins: the latter become more and more important as the temperature is increased. In the high-temperature phase, spins are oriented randomly, leading to a more uniform distribution of 1-cycle sizes. Using persistent homology we are able to easily identify the magnetization as the order parameter, as is well known.

\begin{figure}[t]
	\centering
	\includegraphics[width=0.9\textwidth]{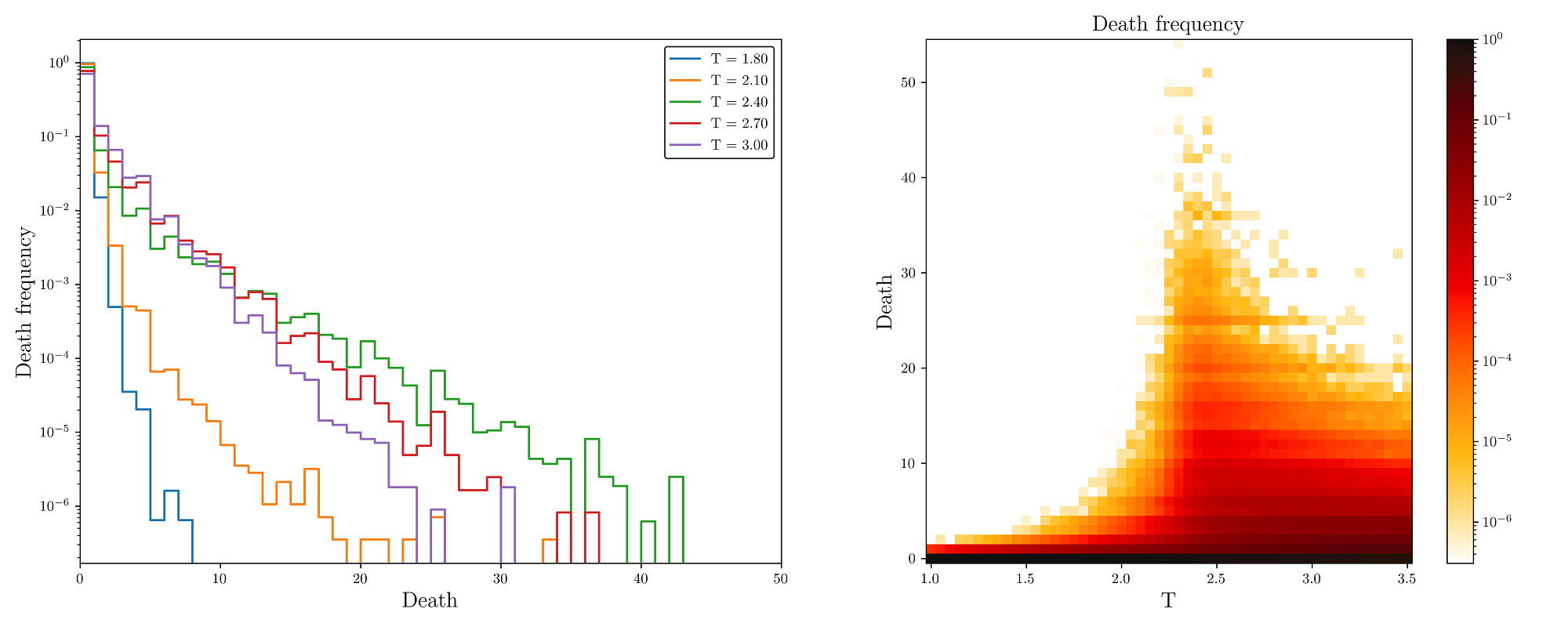}
	\caption{Ising model death distributions. The slight horizontal stripes in the figure on the right (e.g.~at $\mathrm{death}=25$) are symptomatic of the underlying lattice.}
	\label{fig:Ising_deaths}
\end{figure}

\subsubsection{Multiscale behavior near criticality}
Since persistent homology contains multiscale information about a spin configuration it seems reasonable that one should be able to probe a model's approach towards scale-invariance via critical exponents.
Indeed, we are able to see aspects of scale-invariance appearing at criticality by looking at statistics derived from the persistence diagram. One-dimensional statistics such as the Betti numbers, births and deaths can be found by counting points in different regions of the persistence diagrams. In this way we may compute the 1-cycle death probability density, $\mathrm{D}_T(d)$, at each temperature, which quantifies the distribution of feature sizes in the spins. In Figure~\ref{fig:Ising_deaths} we see that deaths are exponentially distributed with a long tail forming at criticality, indicative of a diverging correlation length and the emergence of power-law behavior.

To be more quantitative, we may fit each $\mathrm{D}_T(d)$ to a function of the form
\begin{equation}
	\mathrm{D}_T(d) = A\,d^{-\mu}e^{-d/a_\text{death}} \,.
\end{equation}
Here $d$ is the filtration parameter at the death scale of a cycle, and $A$ is a numerical constant. There are two critical exponents to be extracted: $\mu$ gives the power-law behavior at criticality, while the correlation area $a_\text{death}$ diverges at criticality according to $a_\text{death}\sim|T-T_\text{c}|^{-\nu_\text{death}}$. We are limited by the IR cutoff of the model, namely the finite area of the lattice, but we may still estimate these exponents. As a consistency check, we ask how these might be related to previously studied critical exponents. Using scaling arguments, one can show that at criticality the proportion of clusters of $k$ aligned spins goes as
\begin{equation}
	\mathrm{P}(\text{cluster of size }k) \sim k^{-\tau} \,,
\end{equation}
where the critical exponent is $\tau\approx 2.032$~\cite{1983JPhA.16.1721B,Toral:1987ba}. The function $\mathrm{D}_T(d)$ is not directly measuring the size of clusters, since the death of a 1-cycle around an island of spins is influenced nontrivially by the shape and ``nesting'' of clusters. Nevertheless, it seems reasonable to expect that at criticality the distribution of 1-cycle deaths should follow a similar power-law distribution. Recall that the value of the filtration parameter at the death of a 1-cycle is the \emph{area} of the disks placed on each point in the point cloud, and so roughly corresponds to the number, $k$, of spins enclosed by the 1-cycle.

\begin{figure}[t]
	\centering
	\includegraphics[width=0.9\textwidth]{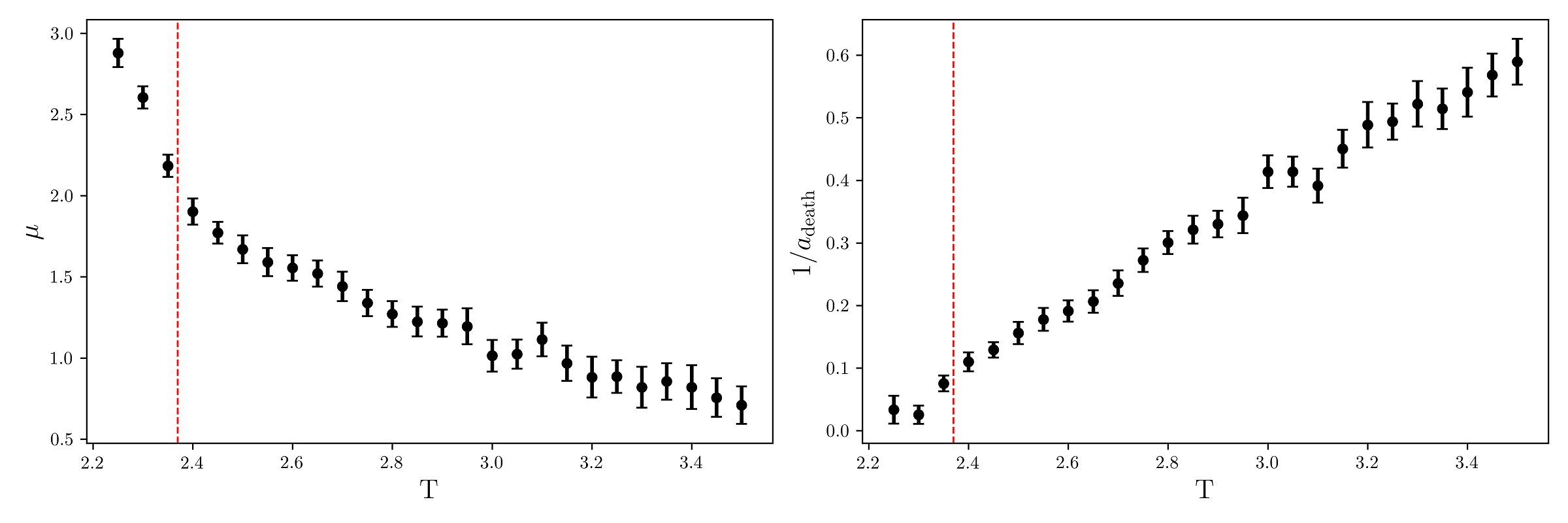}
	\caption{$\mu$ and $a_\text{death}$ for Ising death distributions. The red dashed lines indicates the previously estimated critical temperature $T\approx2.37$. Error estimates are derived from fitting multiple simulations.}
	\label{fig:death_fit_params}
\end{figure}

The fit parameters $\mu$ and $a_\text{death}$ are shown in Figure~\ref{fig:death_fit_params}, where we see clearly the diverging correlation area as criticality is approached from above. The value of $\mu$ at our previously estimated critical temperature, $T\approx 2.37$, is consistent with $\mu\approx \tau=2.032$ as anticipated above, although a more detailed study would be needed to determine the value of $\mu$ more exactly. We see also the linear behavior of $a_\text{death}^{-1}$ with temperature, indicating $\nu_\text{death}\approx 1$. That this is the same degree of divergence as the correlation length of the spin-spin correlation function $\left<s(0)s(r)\right>\sim e^{-r/\xi},~\xi\sim |T-T_c|^{-1}$ can be understood by the following \emph{rough} argument.

The death of a 1-cycle in the $\alpha$-filtration roughly corresponds to the \emph{area} of the cluster of spins that it encloses and $\mathrm{D}_T(d)$ roughly corresponds to the probability that a contiguous region of spins with area $d$ is aligned. Consider for simplicity looking to estimate the probability that a disk of spins with radius $R$ are all aligned. At infinite temperature where the spins are randomly aligned, this probability would simply be $2^{-\pi R^2/\ell^2}$, where $\ell$ is the lattice spacing. If we suppose that $\mathrm{P}_\text{disk}(R)\sim e^{-R^2/a}$ for some ``correlation area'' $a$ even at finite temperature, then how is $a$ related to $\xi$ as defined by $\langle s(R)s(0)\rangle\sim e^{-R/\xi}$ in the disordered phase? To estimate $\mathrm{P}_\text{disk}(R+\ell)\sim e^{-(R+\ell)^2/a}\approx e^{-R^2/a-2\ell R/a}$, imagine asking that a circle of $\sim R$ spins all be aligned with the disk of (aligned) spins of radius $R$ that they encircle. For simplicity, we ignore conditional aspects of the probability and subleading terms. This should then take the form $e^{-R^2/a}e^{-\#R/\xi}$, from which we conclude that $a\sim \ell\xi$: in particular, $a\sim\xi\sim|T-T_\text{c}|^{-\nu}$ with the same critical exponent as the critical temperature is approached from above.

It would be interesting to further understand the relationship between the persistence critical exponents we defined and those typically studied.

\subsection{Square-ice model}\label{sec:sqIce}

\begin{figure}[!t]
	\centering
	\begin{subfigure}{0.9\textwidth}
		\centering
		\includegraphics[width=\textwidth]{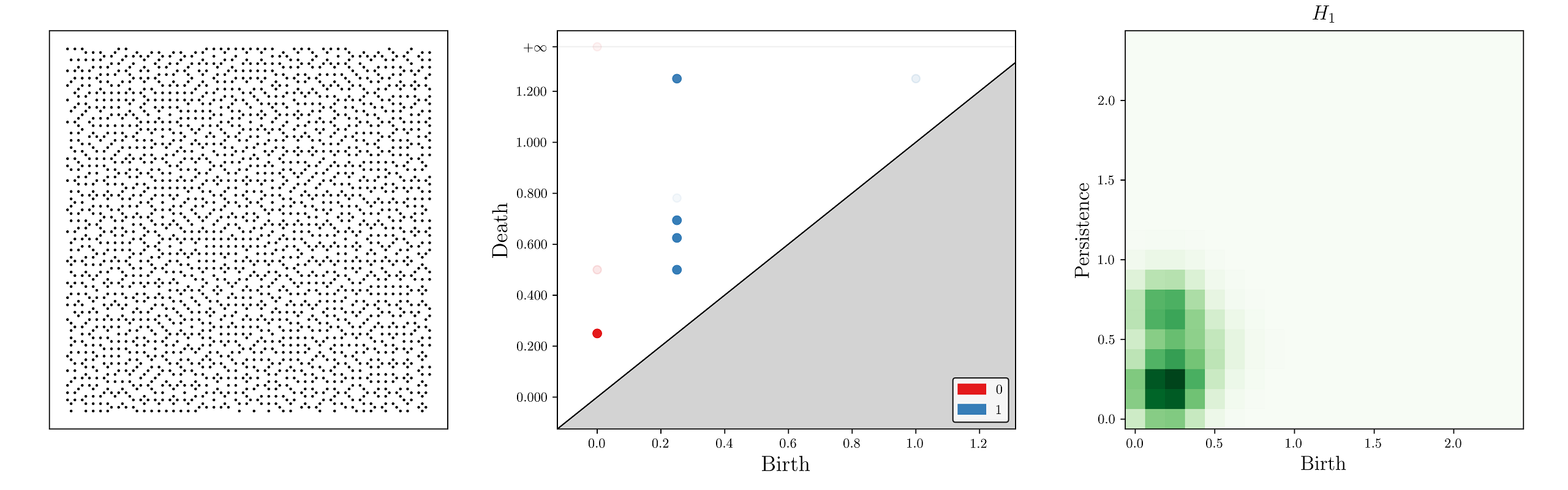}
		\includegraphics[width=\textwidth]{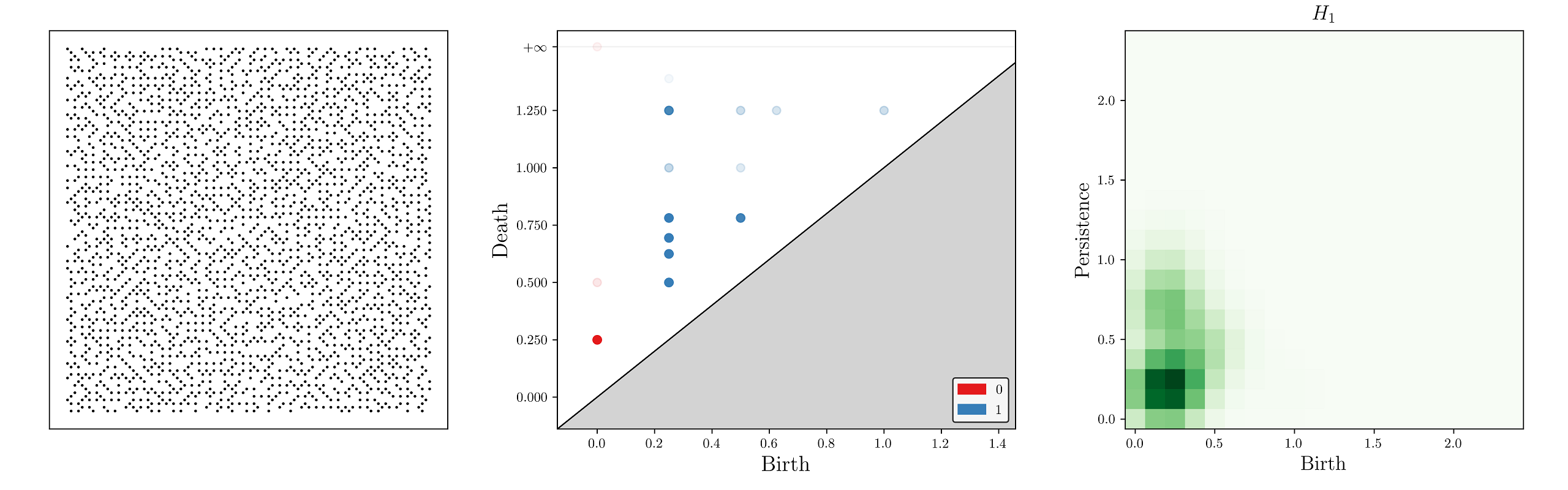}
		\subcaption{Sample spin configuration, persistence diagram and persistence image for $T=0.5$ (top) and $T=4.0$ (bottom).}
		\label{fig:SquareIce_PI}
	\end{subfigure}
	\begin{subfigure}{0.9\textwidth}
		\centering
		\includegraphics[width=\textwidth]{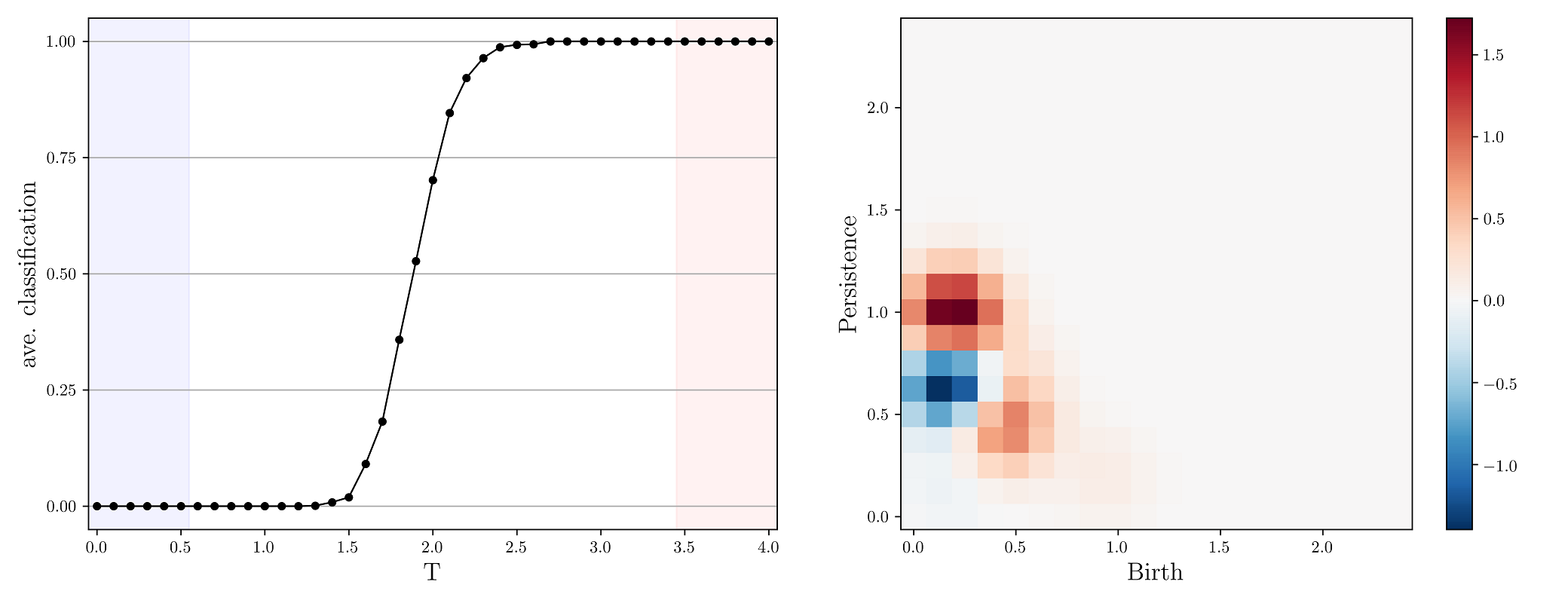}
		\subcaption{Average classification of testing data and learned logistic regression coefficients for the square-ice model. The training data have temperatures in the highlighted regions. In the regression coefficients, blue regions are more populated in the low-temperature phase and red regions are more populated in the high-temperature phase.}
		\label{fig:SquareIce_log_reg}
	\end{subfigure}
	\caption{Square-ice persistence data and phase classification.}
\end{figure}

The square-ice model places spins, $s_i\in\{{-1},1\}$, on the edges rather than vertices of a square lattice and is governed by the local Hamiltonian
\begin{equation}
	H_\text{SI} = \sum_{v\in\Lambda}\Big(\sum_{i:v}s_i\Big)^2 \,,
\end{equation}
where $i:v$ denotes those spins on edges adjacent to the vertex $v$. In contrast to the Ising model there is no spontaneous magnetization at low temperatures. Rather, the ground state is highly degenerate: any configuration with exactly two up and two down spins adjacent to every vertex has zero energy. This leads to frustration in the low-energy dynamics, as adjacent vertices $v$ compete to minimize $\left(\sum_{i:v}s_i\right)^2$. This competition takes place at small scales, so that
many 1-cycles die very quickly in the filtration. Nevertheless, we are still able to identify a shift in the distribution of $p$-cycle births and deaths and reliably classify samples into two phases. In this case, the frustration introduces a particular length scale to the topological features in the low-temperature phase, while the distribution of sizes in the high-temperature phase is less restricted.

\subsubsection{Logistic regression and order parameter}
We generate 1000 sample spin configurations for a $50\times 50$ lattice with $5000$ spins at temperatures $T\in\{0.0,0.1,\ldots,4.0\}$. Each sample gives a persistence image with a weight $\log{(1+p)}$, such as that shown in Figure~\ref{fig:SquareIce_PI}. Again training a logistic regression only on those persistence images with extreme temperatures (Figure~\ref{fig:SquareIce_log_reg}), we find an estimate of $T\approx 1.9$ for the critical temperature. From the logistic regression coefficients on the right-hand side of Figure~\ref{fig:SquareIce_log_reg} we see that as the temperature increases there is a tendency for 1-cycles to be born later or to be longer-lived. Both are indicative of a changing local structure in the spin configurations. In the low-temperature phase, it is energetically beneficial for neighboring vertices to coordinate, resulting in an regular patterns of alternating up and down spins. This regularity forces 1-cycles to live at smaller scales than in the high-temperature phase.

\subsection{XY model}\label{sec:xy}

\begin{figure}[!t]
	\centering
	\begin{subfigure}{0.9\textwidth}
		\centering
		\includegraphics[width=\textwidth]{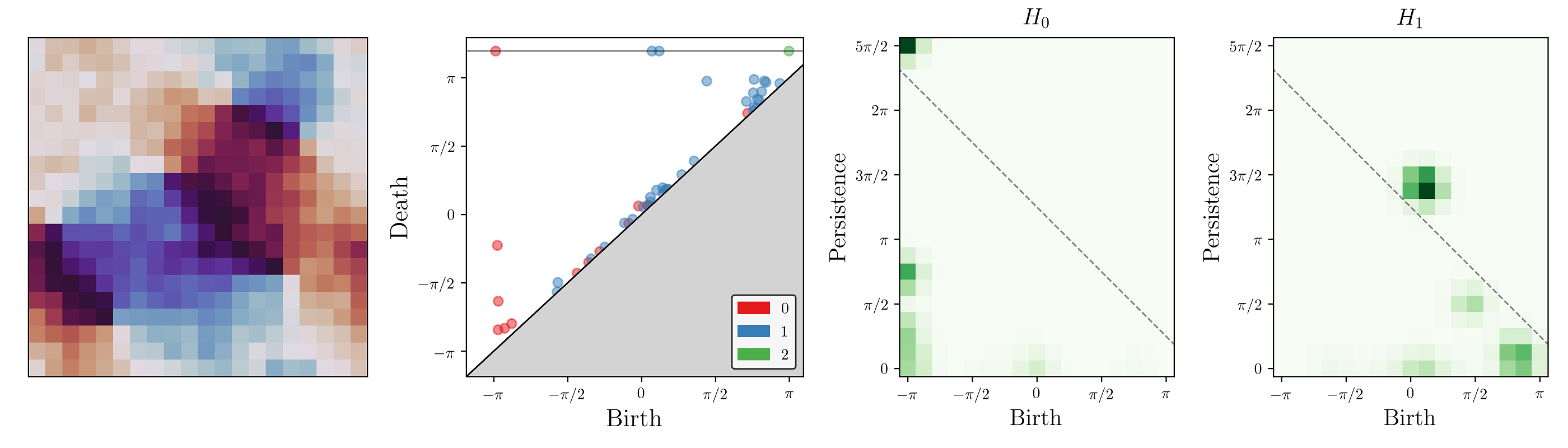}
		\includegraphics[width=\textwidth]{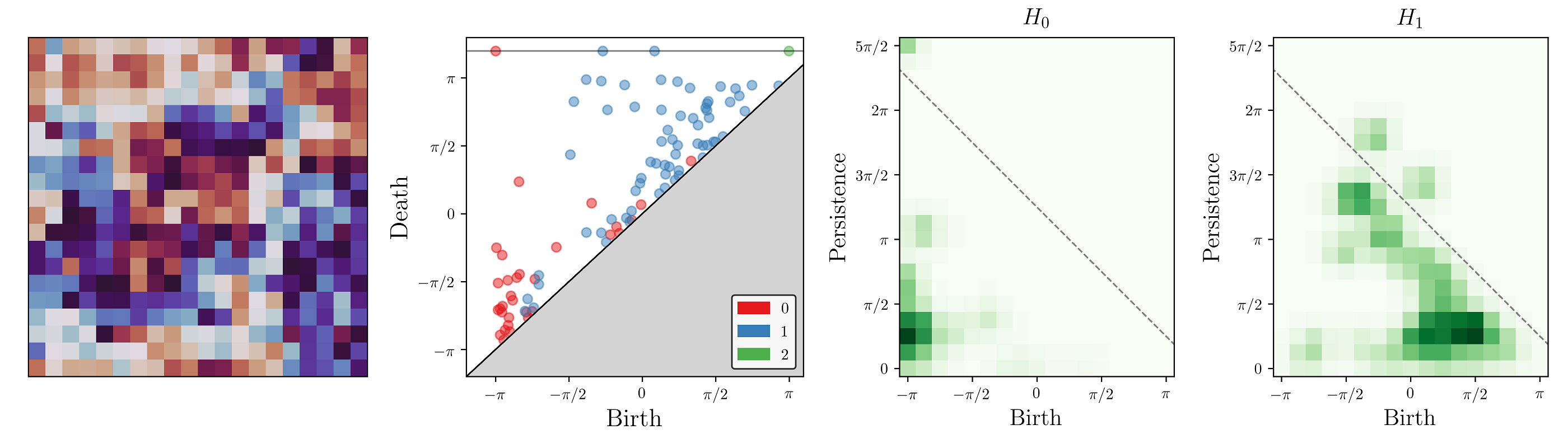}
		\subcaption{Sample spin configuration, persistence diagram and persistence image for $T=0.10$ (top) and $T=1.40$ (bottom).}
		\label{fig:XY_PI}
	\end{subfigure}
	\begin{subfigure}{0.9\textwidth}
		\centering
		\includegraphics[width=\textwidth]{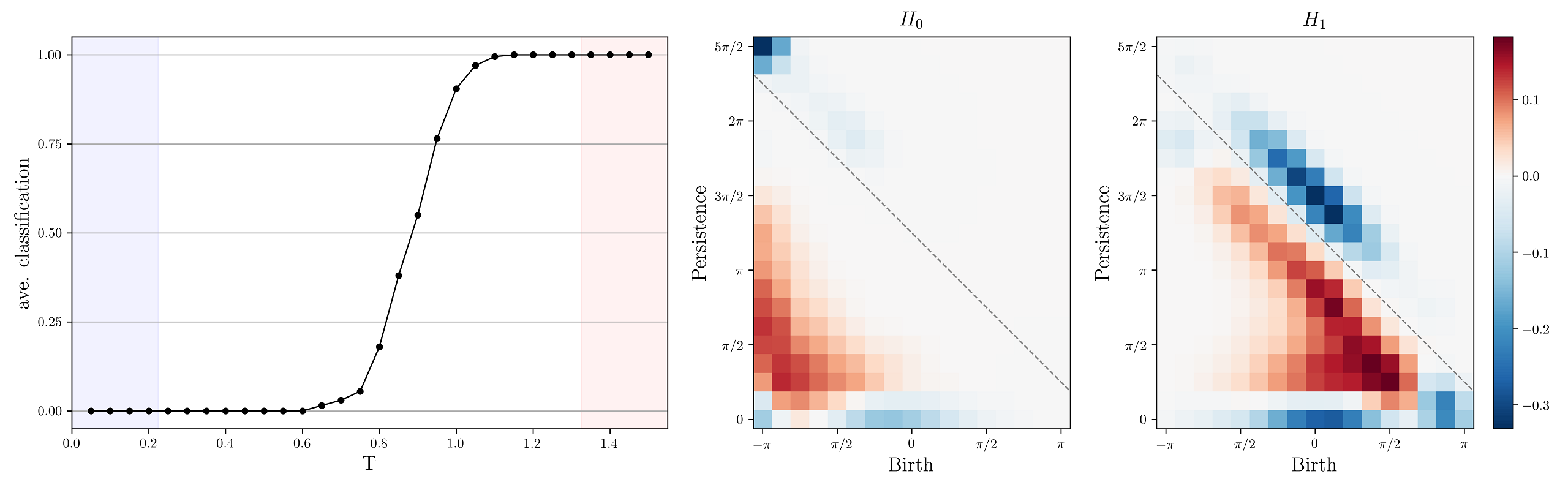}
		\subcaption{Average classification of testing data and learned logistic regression coefficients for the XY model. The training data have temperatures in the highlighted regions. In the regression coefficients, blue regions are more populated in the low-temperature phase while red regions are more populated in the high-temperature phase.}
		\label{fig:XY_log_reg}
	\end{subfigure}
	\caption{XY persistence data and phase classification. The dashed lines visually separate the infinite persistence 0- and 1-cycles from those which have finite death.}
\end{figure}

The XY model is a continuous-spin generalization of the Ising model. At each site of the square lattice spins take values in $S^1$ and are governed by
\begin{equation}
	H_\text{XY} = -\sum_{\langle i,j \rangle}\cos{(\theta_i-\theta_j)} \,.
\end{equation}
There is a well-known KT phase transition at $T_\text{XY}\approx 0.892$ (see~\cite{PhysRevB.37.5986,Hasenbusch:2005xm}, among others). This is an infinite-order phase transition where at low temperatures there are bound vortex-antivortex pairs while at high temperatures free vortices proliferate and spins are randomly oriented.

\subsubsection{Logistic regression and order parameter}
With continuous spins each spin configuration implicitly contains much more information about the underlying dynamics. For temperatures $T\in\{0.05,0.10,\ldots,1.50\}$ we generate 200 sample spin configurations on a $20\times20$ lattice with 400 spins. Persistence images are created for each sample, as in Figure~\ref{fig:XY_PI}. The zeroth homology, in contrast to the $\alpha$-complexes used for discrete spins, is very rich for the cubical complexes and we include both $H_0$ and $H_1$ persistence data in the persistence images. There is always a single 0-cycle and two 1-cycles which never die: these correspond to the $p$-cycles of the torus on which the lattice lives. We distinguish these immortal $p$-cycles from those cycles with late deaths ($d\approx \pi$) by giving the former a death of $d=\frac{5\pi}{2}$ by hand. Omitting these infinite persistence ``torus cycles'' results in a comparable phase classification. 

\begin{figure}[!t]
	\centering
	\includegraphics[width=0.9\textwidth]{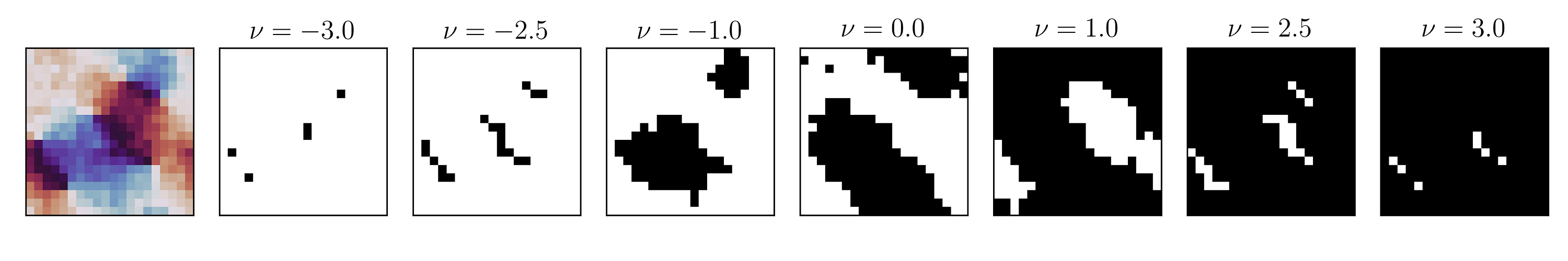}
	\caption{Example spin configuration with three vortex-antivortex pairs, with seven sublevel sets shown. Each vortex-antivortex pair corresponds to a number of 0-cycles which are born very early and 1-cycles which die very late.}
	\label{fig:XY_levelsets}
\end{figure}

Performing a logistic regression of the concatenated $H_0$ and $H_1$ persistence images by training on configurations with temperatures far away from the anticipated transition leads to the classification of Figure~\ref{fig:XY_log_reg}. The critical temperature is estimated as $T_\text{XY}\approx 0.9$. We see that the low-temperature phase is characterized by $p$-cycles on the ``boundary'' of the persistence images. This we can understand in the following way. A (small enough) loop around an isolated vortex has nontrivial winding number, which ensures that there are spins with angles close to both $-\pi$ and $\pi$ if a vortex is present. This explains the strong blue regions in the corners of the logistic regression coefficients: for $\nu\approx -\pi$ a number of 0-cycles are born very early for each vortex and antivortex, giving the lower-left corner of the $H_0$ coefficients. One of these 0-cycle lives forever, giving the upper-left corner of the $H_0$ coefficients. In addition, there are 1-cycles which are born close to $\nu\approx\pi$, again corresponding to the extreme angles associated with the (anti)vortices. See Figure~\ref{fig:XY_levelsets} for an example of this interpretation in practice. When vortex-antivortex pairs happen to not be present at low temperatures, then all of the spins are aligned close to $\theta=0$, giving the short-lived features centered around a birth of zero along the bottom edges of both the $H_0$ and $H_1$ coefficients.

\begin{figure}[!t]
	\centering
	\includegraphics[width=0.9\textwidth]{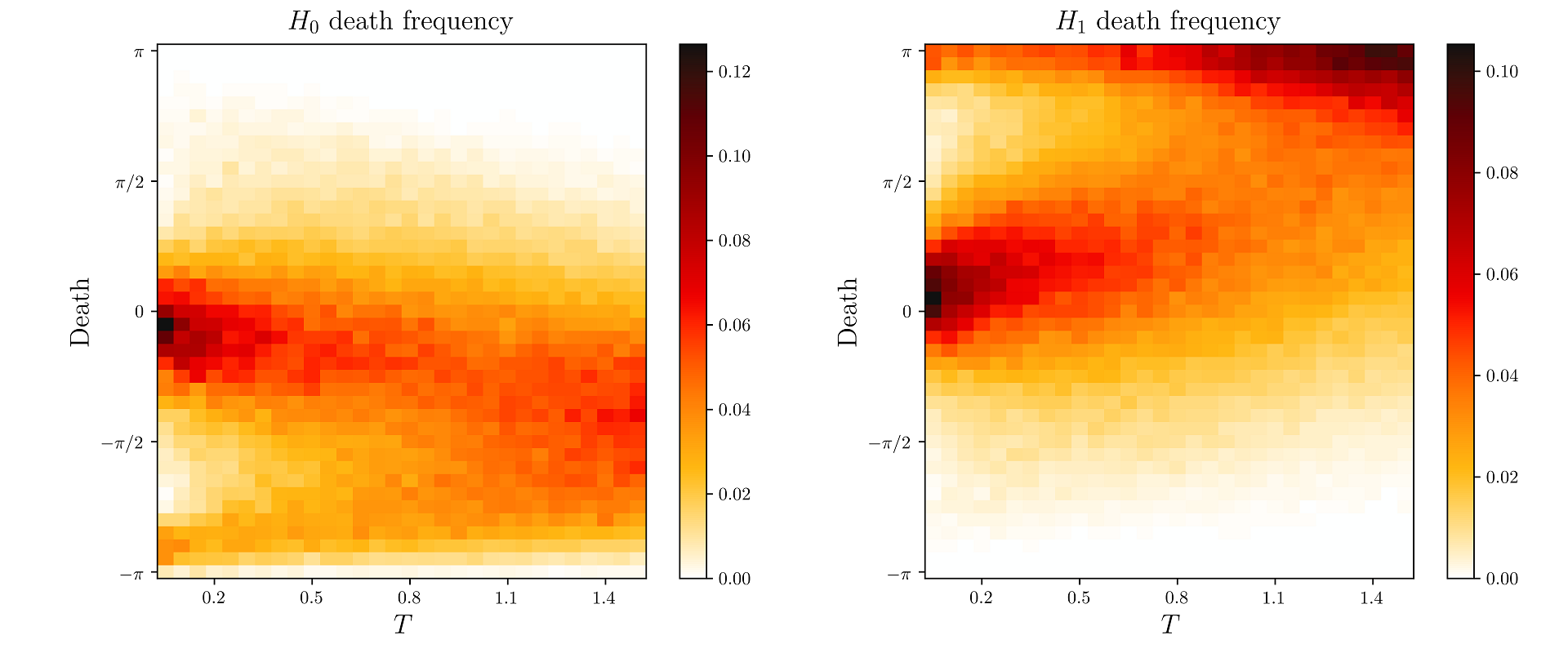}
	\caption{XY 0- and 1-cycle death distributions.}
	\label{fig:XY_deaths}
\end{figure}

\begin{figure}[!t]
	\centering
	\includegraphics[width=\textwidth]{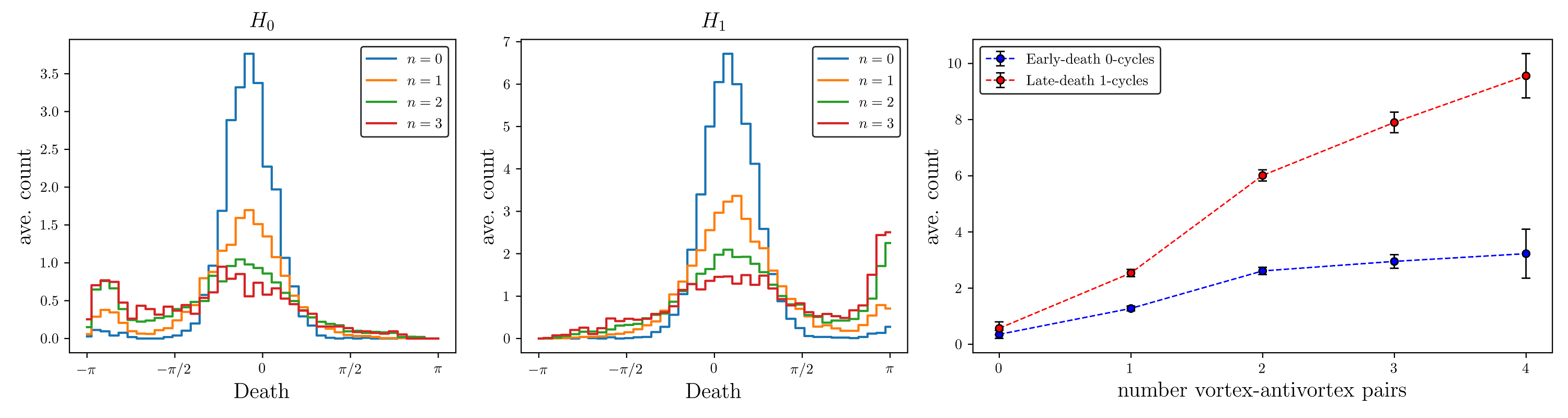}
	\caption{Average counts of 0- and 1-cycle deaths for fixed number of vortex-antivortex pairs at low temperatures ($T\leq 0.20$). (Right) The average counts of 0-cycles with death $d\leq-\frac{3\pi}{4}$ and 1-cycles with $d\geq\frac{3\pi}{4}$. We see that these features are correlated with the number of vortex-antivortex pairs in the configuration.}
	\label{fig:XY_vortex_counts}
\end{figure}

As before we may consider the distribution of $p$-cycle deaths as a function of temperature. In Figure~\ref{fig:XY_deaths} we see that low-temperatures there are two ``populations'' of both 0- and 1-cycles which merge into one as we pass into the high-temperature phase. This again can be attributed to the presence of vortex-antivortex pairs in the following way. Using the raw spin configurations we may count the number of (anti)vortices simply by looking for nontrivial winding in $2\times2$ blocks of the lattice. This can be compared with the number of 0-cycles with early death (e.g.~$d\leq-\frac{3\pi}{4}$) and the number of 1-cycles with late death (e.g.~$d\geq \frac{3\pi}{4}$). Averaging over samples with temperatures below $0.20$ where the number of vortex-antivortex pairs is reasonably small on the $20\times 20$ lattice leads to Figure~\ref{fig:XY_vortex_counts}. There is a clear correlation between the number of extreme-death $p$-cycles and the number of vortex-antivortex pairs as determined directly from the spins. This topological signature of vortex-antivortex pairs should exist rather generally.

Previous investigations of the XY model and its KT phase transition using neural networks and PCA have faced difficulties in identifying vortices at low temperatures~\cite{PhysRevB.97.045207,PhysRevE.95.062122}. It is worth emphasizing the relative ease with which persistent homology identifies (anti)vortices as an important feature at low temperatures.

\subsection{Fully-frustrated XY model}\label{sec:ffxy}

\begin{figure}[!t]
	\centering
	\begin{subfigure}{0.9\textwidth}
		\centering
		\includegraphics[width=\textwidth]{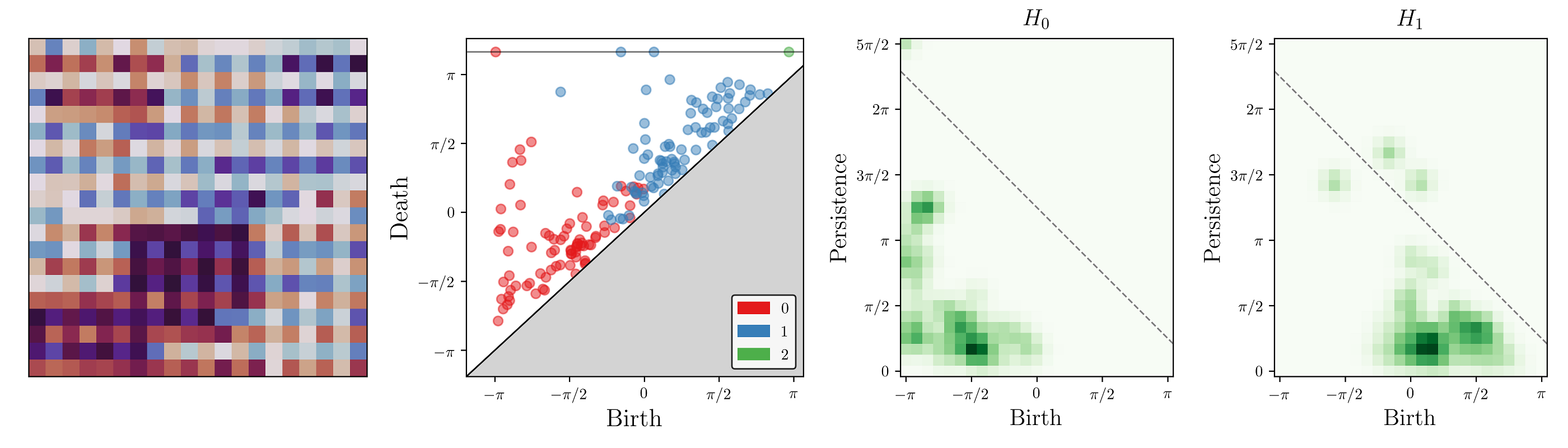}
		\includegraphics[width=\textwidth]{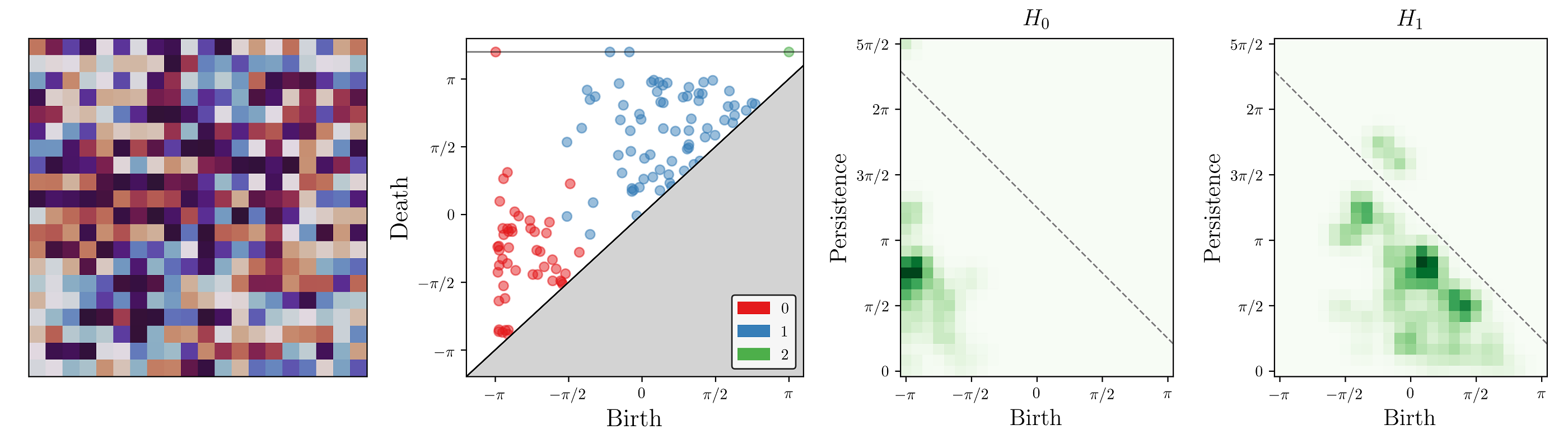}
		\subcaption{Sample spin configuration, persistence diagram and persistence image for $T=0.15$ (top) and $T=1.00$ (bottom).}
		\label{fig:FFXY_PI}
	\end{subfigure}
	\begin{subfigure}{0.9\textwidth}
		\centering
		\includegraphics[width=\textwidth]{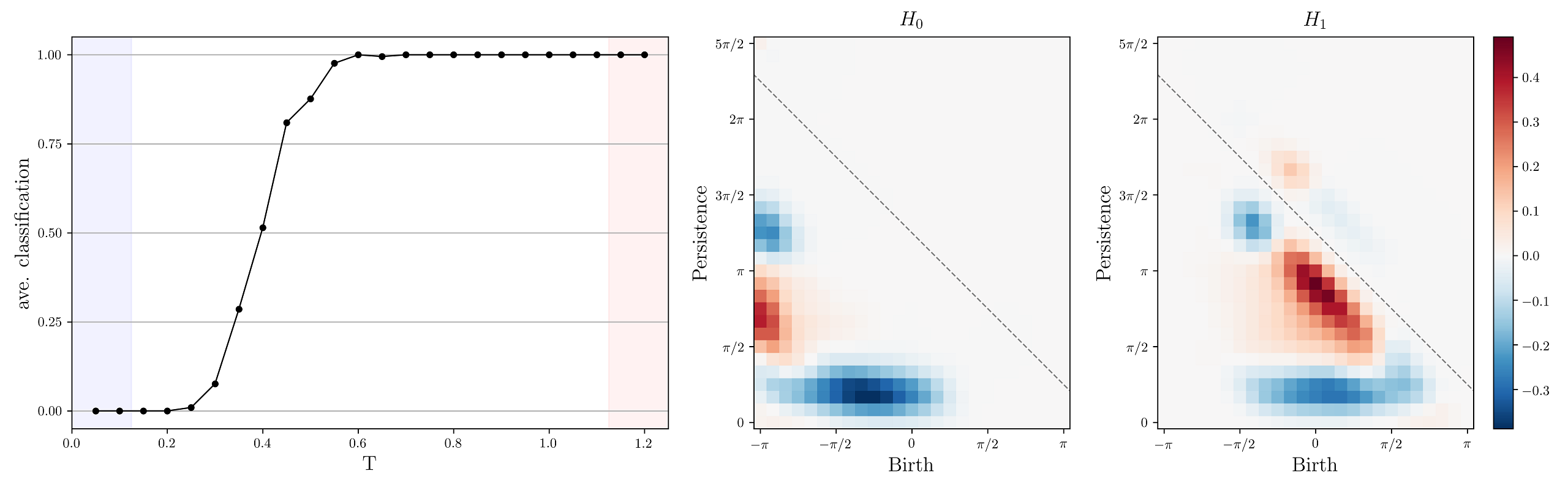}
		\subcaption{Average classification of testing data and learned logistic regression coefficients for the FFXY model. The training data have temperatures in the highlighted regions. In the regression coefficients, blue regions are more populated in the low-temperature phase and red regions are more populated in the high-temperature phase.}
		\label{fig:FFXY_log_reg}
	\end{subfigure}
	\caption{FFXY persistence data and phase classification. The dashed lines visually separate the infinite persistence 0- and 1-cycles from those which have finite death.}
\end{figure}

A frustrated version of the XY model is obtained by changing some of the nearest-neighbor interactions to be antiferromagnetic. One such choice which is fully frustrated is
\begin{equation}
	H_\text{FFXY} = -\sum_{\langle i,j\rangle}J_{ij}\cos{(\theta_i-\theta_j)} \,,
\end{equation}
where $J_{ij}={-1}$ on every other row of horizontal edges and $J_{ij}={+1}$ everywhere else. There are two phase transitions that occur at temperatures which are very close together: a phase transition at $T\approx0.454$ is associated with the loss of $\mathbb{Z}_2$ symmetry, and a phase transition at $T\approx0.446$ is associated with the loss of the $\mathrm{SO}(2)$ rotational symmetry~\cite{Hasenbusch:2005sh}. Because of their proximity we are unable to identify both transitions without an extensive set of simulations.
\subsubsection{Logistic regression and order parameter}
\begin{figure}[!t]
	\centering
	\includegraphics[width=0.9\textwidth]{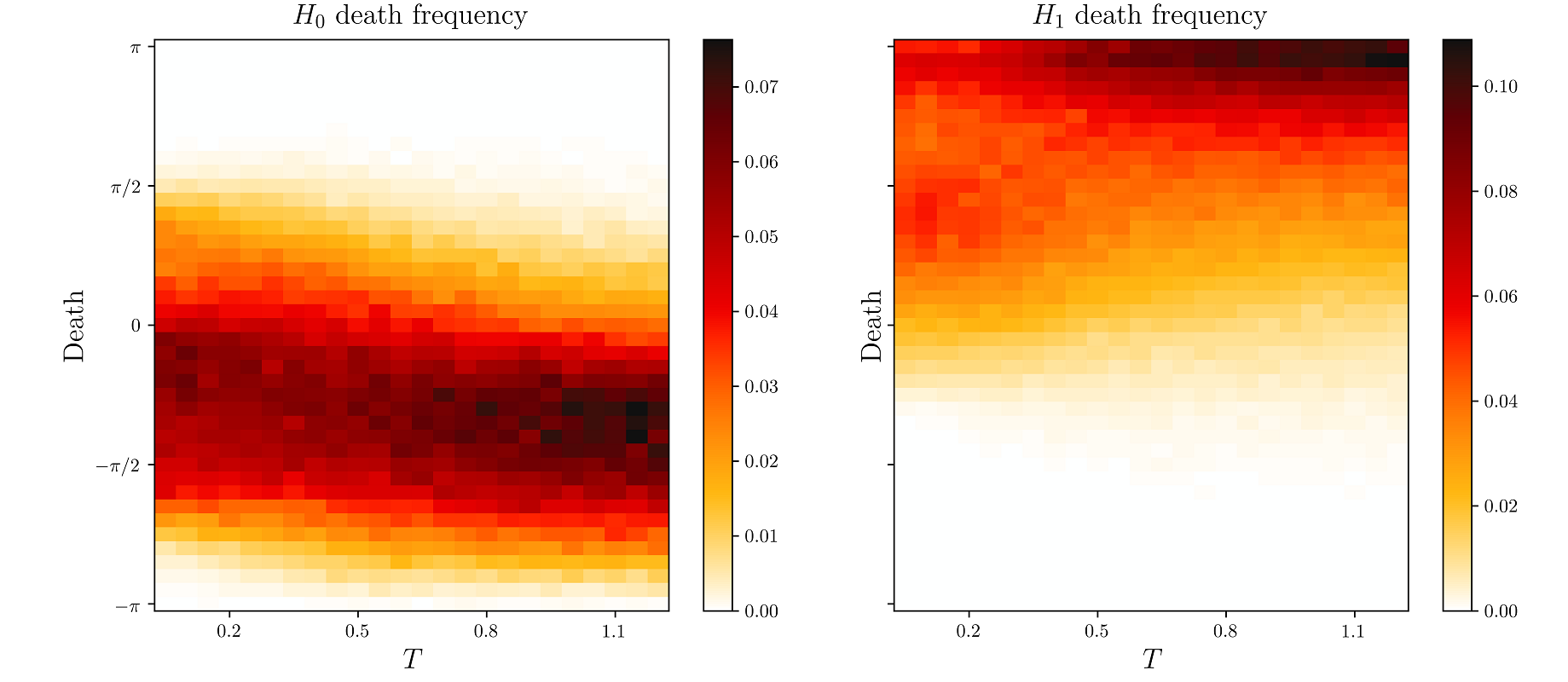}
	\caption{FFXY 0- and 1-cycle death distributions.}
	\label{fig:FFXY_death}
\end{figure}

We generate 200 sample spin configurations on a $20\times 20$ lattice with 400 spins for temperatures $T\in\{0.05,0.10,\ldots 1.20\}$. As before with the XY model, the zeroth homology is quite rich and we include it in the persistence images, such as that in Figure~\ref{fig:FFXY_PI}. Training the logistic regression leads to the classification in Figure~\ref{fig:FFXY_log_reg}, where the critical temperature is estimated as $T_\text{FFXY}\approx0.39$. A more accurate estimation can be achieved by using training data closer to the phase transition. The learned coefficients show a strong tendency for both 0- and 1-cycles to shift to have persistence around $\frac{3\pi}{4}$ in the high-temperature phase. As in the square-ice model, our order parameter probes the small-scale structure of the frustration pattern. In particular, the low-temperature phase exhibits ``pseudo-domains'' where many next-to-nearest neighbors take similar spin values. The alternating structure induced by the antiferromagnetic bands therefore leads to more isolated local minima (i.e.\ 0-cycles in the sublevel filtration) in the low-temperature phase. In the high-temperature phase, most of the local minima are born at $\theta\sim -\pi$, while in the low-temperature phase there are local minima at higher $\theta$ protected by these pseudo-domains. This explains the blue band at the bottom of the $H_0$ logistic regression coefficients. The lack of vortices can be seen from the death distribution as a function of temperature in Figure \ref{fig:FFXY_death}.

In our discussion we have used a sublevel filtration with cubical complex to quantify the homology of continuous-spin configurations. Another approach would be to construct point clouds by taking the locations of spins in a (sub)levelset and using an $\alpha$-filtration. By scanning through levelsets one can capture the topological features of $f:\Lambda\to S^1$ in a different way. For the fully-frustrated XY model this leads to a comparable classification and estimate for the critical temperature.

\section{Discussion}
\label{sec:disc}

In this paper we have explored the use of persistent homology in quantitatively analyzing the phase structure and critical behavior of lattice spin models. While the models we consider can be understood via other means, the use of persistent homology provides an interesting perspective into their statistical properties. Nonlocal features are naturally accounted for in this framework and could prove to be useful in more complex systems as well.

Phase classification using persistence images alone is accomplished successfully for the Ising, square-ice, XY and fully-frustrated XY models, providing a mix of examples with discrete/continuous spins and some with frustration. The resulting trained logistic regressions reveal those regions of the persistence image/diagram which are characteristic of low- and high-temperature phases. This allows for an easily interpretable classification, where (sometimes drastic) shifts in the distributions of $p$-cycle births and deaths are associated with a phase transition. In the case of the XY model, there is a clear correlation between the number of early-death 0-cycles, late-death 1-cycles and the number of bound vortex-antivortex pairs at low temperatures.

The persistence data also display features of critical phenomena. For the Ising model one observes the emergence of power-law behavior in the distribution of 1-cycle deaths as the critical temperature is approached. We are able to estimate two critical exponents associated with this behavior: the correlation area diverges as $a_\text{death}\sim|T-T_\text{c}|^{-\nu_\text{death}}$ with $\nu_\text{death}\approx1$ as expected for the 2D Ising model, and we estimate the critical exponent $\mu$, introduced through $\mathrm{D}_T(d)\sim d^{-\mu}$, to be $\mu\approx 2$, in agreement with expectations from the known power-law behavior of cluster sizes at criticality.

We have demonstrated the quantitative statistical capabilities of persistent homology for relatively simple 2D lattice spin systems. It would interesting to apply these ideas and techniques to more complicated lattice spin models in higher dimensions or with no known order parameter. In more than two dimensions the higher homology groups may serve useful in quantifying nonlocal structures. We leave such work for the future.

\section*{Acknowledgments}

We thank Jeff Schmidt for useful discussions. We also thank the participants of the ``Theoretical Physics for Machine Learning'' and ``Physics $\cap$ ML'' workshops at the Aspen Center for Physics and Microsoft Research, respectively, where partial results of this work were presented in early 2019, for discussion. G.J.L.\ and G.S.\ are supported in part by the DOE grant DE-SC0017647 and the Kellett Award of the University of Wisconsin. The code and data used in our analysis is provided in the following GitHub repository: \href{https://github.com/gloges/TDA-Spin-Models}{\texttt{gloges/TDA-Spin-Models}}.

\bibliography{TDA_spins}

\providecommand{\href}[2]{#2}\begingroup\raggedright\begin{thebibliography}{10}

\bibitem{Wang_2016}
L.~Wang,  {\em Discovering phase transitions with unsupervised learning},
  Physical Review B {\bf 94} (Nov, 2016).

\bibitem{van_Nieuwenburg_2017}
E.~P. van Nieuwenburg, Y.-H. Liu and S.~D. Huber,  {\em Learning phase
  transitions by confusion}, Nature Physics {\bf 13} (Feb, 2017) 435–439.

\bibitem{Wetzel_2017}
S.~J. Wetzel and M.~Scherzer,  {\em Machine learning of explicit order
  parameters: From the Ising model to SU(2) lattice gauge theory}, Physical
  Review B {\bf 96} (Nov, 2017).

\bibitem{Hu_2017}
W.~Hu, R.~R.~P. Singh and R.~T. Scalettar,  {\em Discovering phases, phase
  transitions, and crossovers through unsupervised machine learning: A critical
  examination}, Physical Review E {\bf 95} (Jun, 2017).

\bibitem{Woloshyn:2019oww}
R.~Woloshyn,  {\em {Learning phase transitions: comparing PCA and SVM}},
  \href{http://www.arXiv.org/abs/1905.08220}{{\tt 1905.08220}}.

\bibitem{Giannetti_2019}
C.~Giannetti, B.~Lucini and D.~Vadacchino,  {\em Machine Learning as a
  universal tool for quantitative investigations of phase transitions}, Nuclear
  Physics B {\bf 944} (Jul, 2019) 114639.

\bibitem{Carrasquilla_2017}
J.~Carrasquilla and R.~G. Melko,  {\em Machine learning phases of matter},
  Nature Physics {\bf 13} (Feb, 2017) 431–434.

\bibitem{Ch_ng_2017}
K.~Ch’ng, J.~Carrasquilla, R.~G. Melko and E.~Khatami,  {\em Machine Learning
  Phases of Strongly Correlated Fermions}, Physical Review X {\bf 7} (Aug,
  2017).

\bibitem{Huembeli2017:1710.08382v2}
P.~Huembeli, A.~Dauphin and P.~Wittek,  {\em Identifying Quantum Phase
  Transitions with Adversarial Neural Networks}, Physical Review B {\bf 97} (4,
  2018) [\href{http://www.arXiv.org/abs/Arxiv:1710.08382v2}{{\tt
  Arxiv:1710.08382v2}}].

\bibitem{Tanaka2016:1609.09087v2}
A.~Tanaka and A.~Tomiya,  {\em Detection of phase transition via convolutional
  neural network}, Journal of the Physical Society of Japan {\bf 86} (2016)
  063001 [\href{http://www.arXiv.org/abs/Arxiv:1609.09087v2}{{\tt
  Arxiv:1609.09087v2}}].

\bibitem{Carrasquilla_2020}
J.~Carrasquilla,  {\em Machine learning for quantum matter}, Advances in
  Physics: X {\bf 5} (Jan, 2020) 1797528.

\bibitem{edelsbrunner2000topological}
H.~Edelsbrunner, D.~Letscher and A.~Zomorodian,  {\em Topological persistence
  and simplification}, in {\em Proceedings 41st annual symposium on foundations
  of computer science}, pp.~454--463, IEEE.
\newblock 2000.

\bibitem{zomorodian2005computing}
A.~Zomorodian and G.~Carlsson,  {\em Computing persistent homology}, Discrete
  and Computational Geometry {\bf 33} (2005), no.~2, 249--274.

\bibitem{zomorodian2005topology}
A.~J. Zomorodian, {\em Topology for computing}, vol.~16.
\newblock Cambridge university press, 2005.

\bibitem{edelsbrunner2010computational}
H.~Edelsbrunner and J.~Harer, {\em Computational topology: an introduction}.
\newblock American Mathematical Soc., 2010.

\bibitem{murugan2019introduction}
J.~Murugan and D.~Robertson,  {\em An Introduction to Topological Data Analysis
  for Physicists: From LGM to FRBs}, 2019.

\bibitem{desilva2007}
V.~de~Silva and R.~Ghrist,  {\em Coverage in sensor networks via persistent
  homology}, Algebr. Geom. Topol. {\bf 7} (2007), no.~1, 339--358.

\bibitem{carlsson2008local}
G.~Carlsson, T.~Ishkhanov, V.~De~Silva and A.~Zomorodian,  {\em On the local
  behavior of spaces of natural images}, International journal of computer
  vision {\bf 76} (2008), no.~1, 1--12.

\bibitem{chan2013topology}
J.~M. Chan, G.~Carlsson and R.~Rabadan,  {\em Topology of viral evolution},
  Proceedings of the National Academy of Sciences {\bf 110} (2013), no.~46,
  18566--18571.

\bibitem{Xia_2014}
K.~Xia and G.-W. Wei,  {\em Persistent homology analysis of protein structure,
  flexibility, and folding}, International Journal for Numerical Methods in
  Biomedical Engineering {\bf 30} (Jun, 2014) 814–844.

\bibitem{Gameiro2015}
M.~Gameiro, Y.~Hiraoka, S.~Izumi, M.~Kramar, K.~Mischaikow and V.~Nanda,  {\em
  A topological measurement of protein compressibility}, Japan Journal of
  Industrial and Applied Mathematics {\bf 32} (Mar, 2015) 1--17.

\bibitem{2018arXiv180605167S}
A.~E. {Sizemore}, J.~{Phillips-Cremins}, R.~{Ghrist} and D.~S. {Bassett},  {\em
  {The importance of the whole: topological data analysis for the network
  neuroscientist}}, ArXiv e-prints (June, 2018)
  [\href{http://www.arXiv.org/abs/1806.05167}{{\tt 1806.05167}}].

\bibitem{Cole:2017kve}
A.~Cole and G.~Shiu,  {\em {Persistent Homology and Non-Gaussianity}}, JCAP
  {\bf 1803} (2018), no.~03, 025
[\href{http://www.arXiv.org/abs/1712.08159}{{\tt 1712.08159}}].

\bibitem{Biagetti:2020skr}
M.~Biagetti, A.~Cole and G.~Shiu,  {\em {The Persistence of Large Scale
  Structures I: Primordial non-Gaussianity}},
  \href{http://www.arXiv.org/abs/2009.04819}{{\tt 2009.04819}}.

\bibitem{Cirafici_2016}
M.~Cirafici,  {\em Persistent homology and string vacua}, Journal of High
  Energy Physics {\bf 2016} (Mar, 2016).

\bibitem{Cole:2018emh}
A.~Cole and G.~Shiu,  {\em {Topological Data Analysis for the String
  Landscape}}, JHEP {\bf 03} (2019) 054
  [\href{http://www.arXiv.org/abs/1812.06960}{{\tt 1812.06960}}].

\bibitem{adams2015persistence}
H.~Adams, S.~Chepushtanova, T.~Emerson, E.~Hanson, M.~Kirby, F.~Motta,
  R.~Neville, C.~Peterson, P.~Shipman and L.~Ziegelmeier,  {\em Persistence
  Images: A Stable Vector Representation of Persistent Homology}, 2015.

\bibitem{donato2016persistent}
I.~Donato, M.~Gori, M.~Pettini, G.~Petri, S.~De~Nigris, R.~Franzosi and
  F.~Vaccarino,  {\em Persistent homology analysis of phase transitions},
  Physical Review E {\bf 93} (2016), no.~5, 052138.

\bibitem{2020arXiv200102616S}
D.~{Spitz}, J.~{Berges}, M.~K. {Oberthaler} and A.~{Wienhard},  {\em {Finding
  universal structures in quantum many-body dynamics via persistent homology}},
  arXiv e-prints (Jan., 2020) arXiv:2001.02616
  [\href{http://www.arXiv.org/abs/2001.02616}{{\tt 2001.02616}}].

\bibitem{tran2020topological}
Q.~H. Tran, M.~Chen and Y.~Hasegawa,  {\em Topological Persistence Machine of
  Phase Transitions}, arXiv preprint arXiv:2004.03169 (2020).

\bibitem{olsthoorn2020finding}
B.~Olsthoorn, J.~Hellsvik and A.~V. Balatsky,  {\em Finding hidden order in
  spin models with persistent homology}, 2020.

\bibitem{gudhi:AlphaComplex}
V.~Rouvreau,  {\em Alpha complex}, in {\em {GUDHI} User and Reference Manual}.
\newblock {GUDHI Editorial Board}, {3.3.0}~ed., 2020.

\bibitem{milnor2016morse}
J.~Milnor, {\em Morse theory.(AM-51)}, vol.~51.
\newblock Princeton university press, 2016.

\bibitem{gudhi:CubicalComplex}
P.~Dlotko,  {\em Cubical complex}, in {\em {GUDHI} User and Reference Manual}.
\newblock {GUDHI Editorial Board}, {3.3.0}~ed., 2020.

\bibitem{PhysRev.65.117}
L.~Onsager,  {\em Crystal Statistics. I. A Two-Dimensional Model with an
  Order-Disorder Transition}, Phys. Rev. {\bf 65} (Feb, 1944) 117--149.

\bibitem{1983JPhA.16.1721B}
A.~D. {Bruce} and D.~J. {Wallace},  {\em {Droplet theory in low dimensions:
  Ising systems in zero field}}, Journal of Physics A Mathematical General {\bf
  16} (June, 1983) 1721--1769.

\bibitem{Toral:1987ba}
R.~Toral and C.~Wall,  {\em {Finite-size scaling study of the equilibrium
  cluster distribution of the two-dimensional Ising model}}, J. Phys. {\bf A20}
  (1987)
4949.

\bibitem{PhysRevB.37.5986}
H.~Weber and P.~Minnhagen,  {\em Monte Carlo determination of the critical
  temperature for the two-dimensional XY model}, Phys. Rev. B {\bf 37} (Apr,
  1988) 5986--5989.

\bibitem{Hasenbusch:2005xm}
M.~Hasenbusch,  {\em {The Two dimensional XY model at the transition
  temperature: A High precision Monte Carlo study}}, J. Phys. A {\bf 38} (2005)
  5869--5884 [\href{http://www.arXiv.org/abs/cond-mat/0502556}{{\tt
  cond-mat/0502556}}].

\bibitem{PhysRevB.97.045207}
M.~J.~S. Beach, A.~Golubeva and R.~G. Melko,  {\em Machine learning vortices at
  the Kosterlitz-Thouless transition}, Phys. Rev. B {\bf 97} (Jan, 2018)
  045207.

\bibitem{PhysRevE.95.062122}
W.~Hu, R.~R.~P. Singh and R.~T. Scalettar,  {\em Discovering phases, phase
  transitions, and crossovers through unsupervised machine learning: A critical
  examination}, Phys. Rev. E {\bf 95} (Jun, 2017) 062122.

\bibitem{Hasenbusch:2005sh}
M.~Hasenbusch, A.~Pelissetto and E.~Vicari,  {\em {Multicritical behavior in
  the fully frustrated XY model and related systems}}, J. Stat. Mech. {\bf
  0512} (2005) P12002 [\href{http://www.arXiv.org/abs/cond-mat/0509682}{{\tt
  cond-mat/0509682}}].

\end{thebibliography}\endgroup

\end{document}